\documentclass[12pt]{article}
\usepackage{epsfig}
\usepackage{amsmath}
\usepackage{hhline}
\usepackage{amssymb}
\usepackage{times}
\usepackage{cite}
\usepackage{lineno}
\usepackage{xcolor} 

\usepackage{url}
\usepackage[figuresright]{rotating}
\usepackage{hyperref} 

\begin{document}

\begin{titlepage}


\vspace*{2cm}

\begin{center}
\begin{Large}

{\boldmath \bf
    Production of $\pi^+$ and $K^+$ mesons in argon-nucleus
    interactions at 3.2~AGeV
}

\vspace{0.5cm}

BM@N Collaboration

\end{Large}
\end{center}




{\noindent
S. Afanasiev$^1$, G. Agakishiev$^1$, E. Aleksandrov$^1$, I. Aleksandrov$^1$, P. Alekseev$^3$, K. Alishina$^1$, E. Atkin$^6$, T. Aushev$^5$, V. Babkin$^1$, N. Balashov$^1$
% ...
}

\vspace{-0.1cm}

{\small \noindent
$^1$ Joint Institute for Nuclear Research (JINR), Dubna, Russia

\noindent
$^2$ Institute for Nuclear Research of the RAS (INR RAS), Moscow, Russia

\noindent
$^3$ Kurchatov Institute, NRC, Moscow, Russia

\noindent
$^4$ Lebedev Physical Institute of the Russian Academy of Sciences (LPI RAS), Moscow, Russia

\noindent
$^5$ Moscow Institute of Physics and Technology (MIPT), Moscow, Russia

\noindent
$^6$ National Research Nuclear University MEPhI, Moscow, Russia

\noindent
$^7$ Skobeltsyn Institute of Nuclear Physics, Moscow State University (SINP MSU), Moscow, Russia

\noindent
$^8$ St Petersburg University (SPbU), St Petersburg, Russia

\noindent
$^9$ Plovdiv University “Paisii Hilendarski”, Plovdiv, Bulgaria

\noindent
$^{10}$ Key Laboratory of Nuclear Physics and Ion-Beam Application (MOE), Institute of Modern Physics, Fudan University, Shanghai, China

\noindent
$^{11}$ Institute of Mechanics, Bulgarian Academy of Sciences, Sofia, Bulgaria
}

\vspace{0.5cm}

\begin{abstract}
\noindent 
{First physics results of the BM@N experiment at the Nuclotron/NICA complex are presented on $\pi^+$
and $K^+$ meson production in interactions of an argon beam with fixed targets
of C, Al, Cu, Sn and Pb at 3.2~AGeV. Transverse momentum distributions, rapidity
spectra and multiplicities of $\pi^+$ and $K^+$ mesons are measured. The results
are compared with predictions of theoretical models and with other measurements
at lower energies.}
\end{abstract}

\vspace{1cm}

\end{titlepage}

\section{Introduction}
\label{sect1}

BM@N (Baryonic Matter at Nuclotron) is the first operational experiment at
the Nuclotron/NICA accelerator complex. The Nuclotron will provide beams
of a variety of particles, from protons up to gold ions, with kinetic energy
in the range from 1 to 6~GeV/nucleon for light ions with Z/A ratio
of $\sim0.5$ and up to 4.5~GeV/nucleon for heavy ions with Z/A ratio
of $\sim0.4$.
 At these energies, the nucleon density in the fireball
created in the collisions of a heavy-ion beam with fixed targets is 3-4 times
higher than the nuclear saturation density \cite{Friman}, thus allowing
studying heavy-ion interactions in the regime of high-density baryonic matter~\cite{Cleymans}.

The primary goal of the experiment, complemented by the MPD experiment
that will use the Nuclotron beam in a collider mode, is to constrain the
parameters of the equation of state (EoS) of high-density nuclear matter
and to search for the conjectured critical end point,  the onset of the
deconfinement phase transition and the onset of the chiral symmetry restoration.

In addition, the Nuclotron energies are high enough to study strange
mesons and (multi)-strange hyperons produced in nucleus-nucleus
collisions close to the kinematic threshold~\cite{NICAWhitePaper,BMN_CDR}.
Studies of the excitation function of strange particle production below
and near the kinematic threshold make it possible to distinguish between
hard and soft behavior of the EoS~\cite{Fuchs}.

In the commissioning phase, in a configuration with limited phase-space coverage,
BM@N collected first experimental data with beams of carbon, argon, and krypton
ions \cite{BMN_QM,BMN_SQM}.  This paper presents first results
on  $\pi^+$ and $K^+$ meson production in 3.2 AGeV argon-nucleus interactions.

The paper is organized as follows. Section \ref{sect2} describes the experimental
set-up and Section \ref{sect3} is devoted to details of the event reconstruction.
Section \ref{sect4} describes the evaluation of the $\pi^+$, $K^+$ reconstruction
efficiency. Section \ref{sect5} addresses the evaluation of the cross sections,
 multiplicities and systematic uncertainties. Experimental results on transverse
 momentum distributions, rapidity spectra, and multiplicities of $\pi^+$ and $K^+$ mesons
 are given in Section \ref{sect6}. The BM@N results are compared with predictions of
 theoretical models and with experimental data on medium-sized nucleus-nucleus
 interactions measured at lower energies. Finally, the results are
 summarized in Section \ref{sect7}.

\section{Experimental set-up}
\label{sect2}

\begin{figure}[htb]
\begin{center}
\vspace{1.5cm}
\hspace{-3.8cm} \includegraphics[width=0.028\textwidth,bb=550 0 590 120]{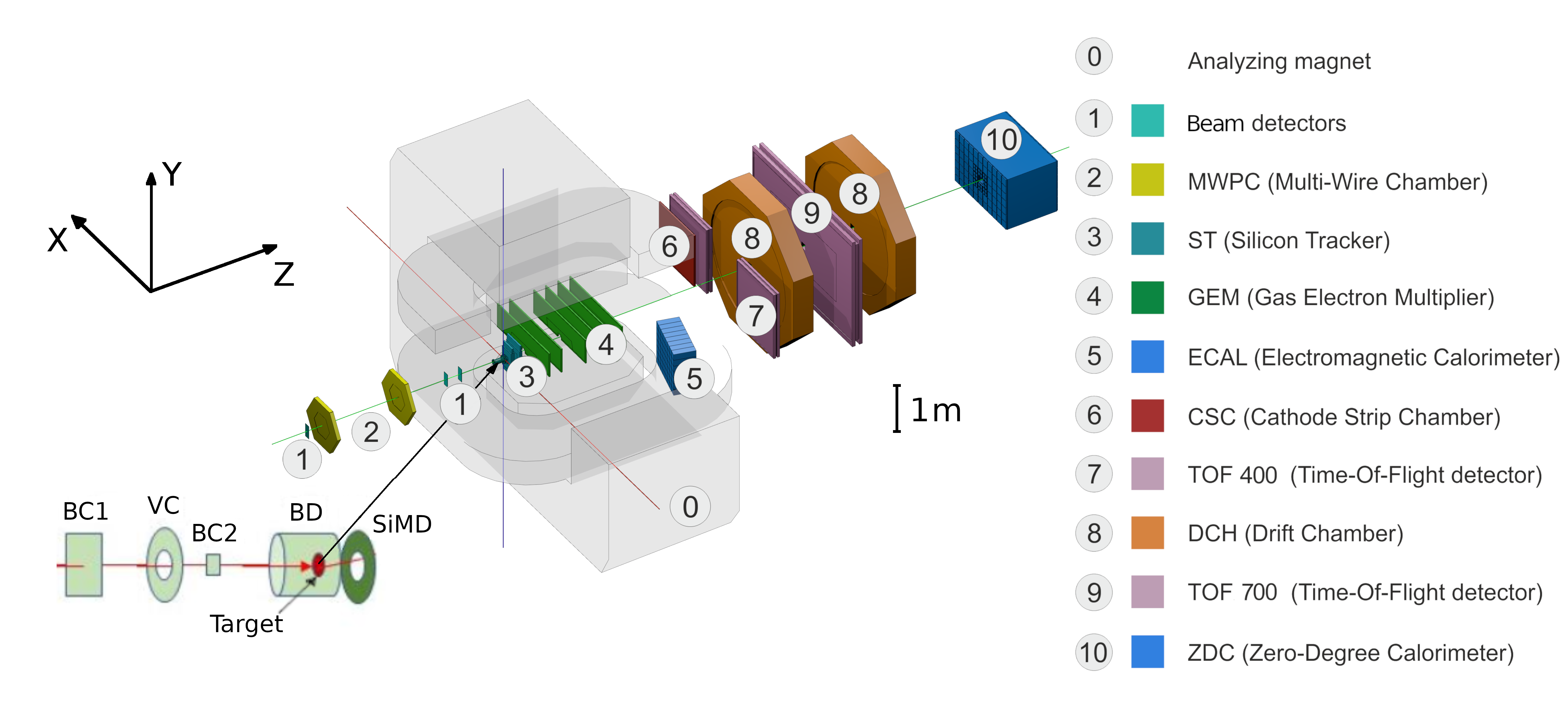}
\end{center}
\vspace{-0.9cm}
\caption {Schematic view of the BM@N setup in the argon beam run.}
\label{BMNsetup}
\end{figure}

The BM@N detector is a forward spectrometer covering the pseudorapidity
range $1.6 \leq \eta \leq 4.4$. A schematic view of the BM@N setup in the
argon-beam run is shown in Fig.~\ref{BMNsetup}. More details of all
components  of the set-up can be found in \cite{BMN_project}.
The spectrometer includes a central tracking system consisting of 3 planes
of forward silicon-strip detectors (ST) and 6 planes of detectors based on
gas electron multipliers (GEM)~\cite{BMN_GEM}. The central tracking system is
 located downstream of the target region inside of a dipole magnet
 with a bending power of about $\approxeq 2.1$Tm and with a gap of 1.05 m
 between the poles . In the measurements reported here, the central tracker
 covered only the upper half of the magnet acceptance.
\vspace{3.5cm}

\begin{figure}[tbh]
\begin{center}
\vspace{0.0cm}
\hspace{-1cm} \includegraphics[width=0.50\textwidth,bb=0 0 620 430]{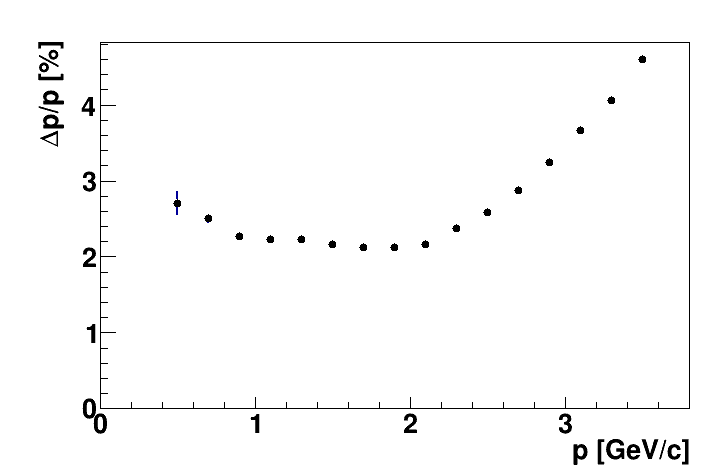}
\end{center}
\vspace{-0.9cm}
\caption {Relative momentum resolution as a function  of the momentum.}
\label{dptop_p}
\end{figure}

Two sets of drift chambers (DCH), a cathode strip chamber (CSC), two sets
of time-of-flight detectors (ToF), and a zero-degree calorimeter (ZDC)
are located downstream of the dipole magnet.  The tracking system measures
the momentum $\rm {p}$ of charged particles with a relative uncertainty
that varies from 2.5\% at a momentum of 0.5 GeV/c to 4.5\% at 3.5 GeV/c as
shown in Fig.~\ref{dptop_p}. The time resolutions of the ToF-400 and
ToF-700 systems are 84 ps and 115 ps, respectively~\cite{KA_2022}.

 Two beam counters (BC1, BC2), a veto counter (VC), a barrel detector (BD),
 and a silicon multiplicity detector (SiMD) were used for event triggering
 and for measurement of the incoming beam ions. The BC2 counter provided
 also the start time T0 for the time of flight measurement. The BD detector
 consists of 40 azimuthal scintillating strips arranged around the target,
 and the SiMD detector consists of 60 azimuthal silicon segments situated behind the target.

To count the number of beam ions that passed through the target, a logical beam trigger
BT = BC1$\land\overline{\rm{VC}}\land$BC2 was used.
The following logic conditions were applied to generate the trigger signal: 1) BT$\land$(BD$\ge\rm{3,4}$);
2) BT$\land$(SiMD$\ge\rm{3,4}$); 3) BT$\land$(BD$\ge\rm{2}$)$\land$(SiMD$\ge\rm{3}$).
The trigger conditions were varied to find the optimal ratio between the event
rate and the trigger efficiency for each target.
Trigger condition 1 was applied for 60\% of the data collected with the carbon target.
This trigger fraction was continuously reduced with the atomic weight of the target
down to 26\% for the Pb target. The fraction of data collected
 with trigger condition 2 was increased from 6\% for the carbon target
 up to 34\% for the Pb target.  The rest of the data were collected with trigger condition 3.
The analysis presented here used the data from the forward silicon detectors, GEM detectors, outer drift chambers,
cathode strip chamber, and the two sets of the time-of-flight detectors
ToF-400 \cite{BMN_ToF400} and ToF-700 \cite{BMN_ToF700}.
Data were collected with an argon beam intensity of a few 10$^5$ ions
per spill and a spill duration of 2-2.5 sec. The kinetic energy of the beam was 3.2 AGeV with the spread of about 1\%. A set of solid targets of various
materials (C, Al, Cu, Sn, Pb) with a relative interaction length of 3\% was used.
The experimental data correspond to an integrated luminosity of 7.8 {$\rm {\mu b^{-1}}$}
collected with the different targets:
2.1 {$\rm {\mu b^{-1}}$} (C), 2.3 {$\rm {\mu b^{-1}}$} (Al),
1.8 {$\rm {\mu b^{-1}}$} (Cu), 1.1 {$\rm {\mu b^{-1}}$} (Sn), 0.5 {$\rm {\mu b^{-1}}$} (Pb).
A total of 16.3M argon-nucleus collisions at 3.2 AGeV were reconstructed.

\section{Event reconstruction}
\label{sect3}

Track reconstruction in the central tracker is based on a
``cellular automaton'' approach~\cite{Kisel} implementing a constrained combinatorial search of
track candidates with their subsequent fitting by a Kalman filter to determine the track parameters.
These tracks are used to reconstruct primary and secondary vertices as well as global tracks by
extrapolation and matching to hits in the downstream detectors (CSC, DCH and ToF).

\begin{figure}[htb]
\begin{center}
\vspace{-0.6cm}
\includegraphics[width=0.5\textwidth,bb=0 0 620 529]{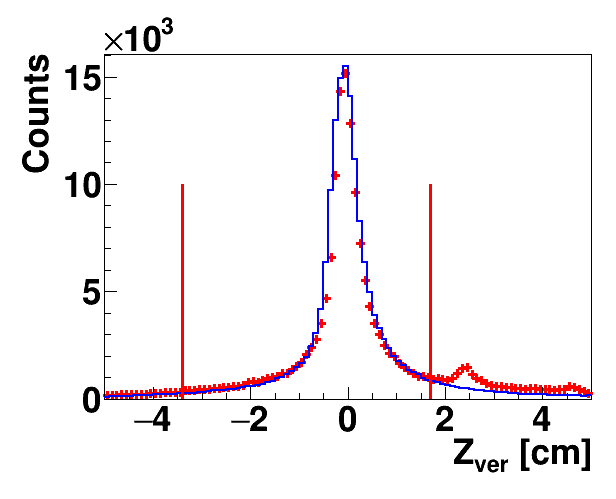}
\end{center}
\vspace{-0.9cm}
\caption {Distribution of the primary vertices along the Z axis for data (red crosses)
          and simulated events (blue histogram). The vertical lines limit the Z region
          accepted for the data analysis.}
\label{zpv_c}
\end{figure}

The primary collision vertex position (PV) is measured with a resolution of 2.4 mm in the X-Y plane
perpendicular to the beam direction and 3 mm in the beam direction at the target position.
The distribution of the primary vertices along the beam direction ($Z_{ver}$) for experimental
data and Monte Carlo events is shown in Fig.~\ref{zpv_c}.

Charged mesons ($\pi^+$ and $K^+$) are identified using the time of flight $\Delta t$
measured between T0 and the ToF detectors, the length of the trajectory $\Delta l$ and
the momentum $p$ reconstructed in the
central tracker. Then the squared mass $M^2$ of a particle is calculated by the
formula: $M^2 = p^2((\Delta t c/\Delta l)^2 - 1)$, where $c$ is the speed of light.

\begin{figure}[tbh]
\begin{center}
\vspace{-0.5cm}
\hspace{1.0cm} (a) \hspace{5.5cm} (b)

\vspace{1.0cm}
\includegraphics[width=0.46\textwidth,bb=0 0 700 350]{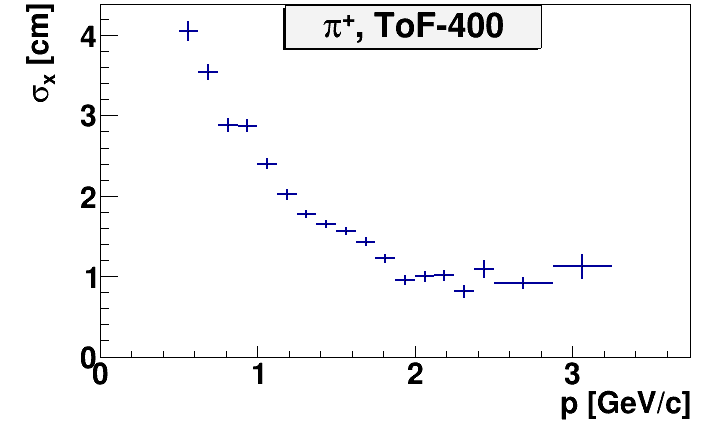}
\includegraphics[width=0.46\textwidth,bb=0 0 700 350]{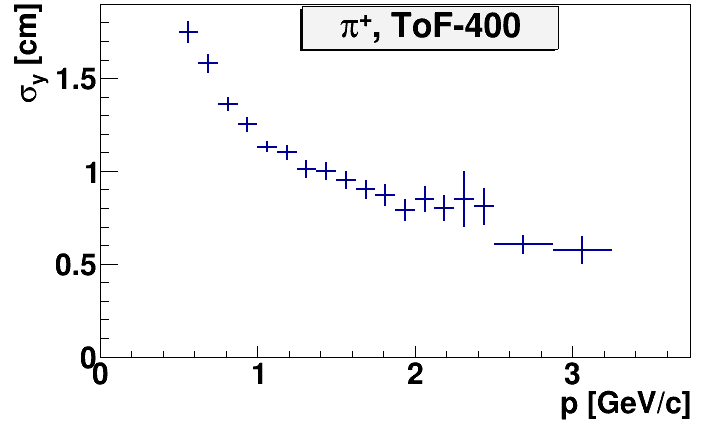}
 \end{center}
 \vspace{-0.4cm}
 \caption {$\sigma_x$ (a) and $\sigma_y$ (b)  of the Gaussian fit of the ToF-400 hit residuals with respect to positively charged pions depending on the particle momentum.}
 \label{dxdyvsp_tof400_data}
\end{figure}

Candidates of $\pi^+$ and $K^+$  must originate from the primary vertex and match
hits in the CSC and ToF-400 or in the DCH and ToF-700 detectors.
The following criteria are required for selecting $\pi^+$ and $K^+$ meson candidates:
\begin{itemize}
\item Each track has at least 4 hits in the GEM detectors (6 detectors in
total)~\cite{BMN_GEM}. Hits in the forward silicon detectors are used to
reconstruct the track, but no requirements are applied to the number of hits;

\item Tracks originate from the primary vertex. The deviation of the reconstructed
vertex from the target position along the beam direction is limited
to -3.4 cm $< Z_{\mathrm {ver}} - Z_0 <$ 1.7 cm, where $Z_0$ is the target position.
The upper limit corresponds to  $\sim 5.7\sigma$ of the $Z_{ver}$ spread and cuts
off interactions with the trigger detector located 3 cm behind the target
(see Fig. \ref{zpv_c}). The two vertical lines in the figure limit the region of
the Z coordinates accepted for the data analysis for all the targets.
The beam interaction rate with the trigger detector is well below 1\% and
was not simulated since it does not affect the precision in Monte Carlo simulation.

\item Distance from a track to the primary vertex in the X-Y plane at Z$_{\mathrm{ver}}${(DCA)}
is required to be less than 1~cm, which corresponds to 4$\sigma$ of the vertex resolution in the X-Y plane;

\item Momentum range of positively charged particles $p>0.5$ GeV/c and $p>0.7$ GeV/c
is limited by the acceptance of the ToF-400 and ToF-700 detectors, respectively;

\item Distance of extrapolated tracks to the CSC (DCH) hits as well as to the
ToF-400 (ToF-700) hits should be within $\pm 2.5\sigma$ of the momentum dependent
hit-track residual distributions as shown in Fig.~\ref{dxdyvsp_tof400_data} for the ToF-400 system.
\end{itemize}

\begin{figure}[tbh]
\begin{center}
\vspace{-0.2cm}
\hspace{0.5cm} (a) \hspace{7cm} (b)
\includegraphics[width=0.49\textwidth]{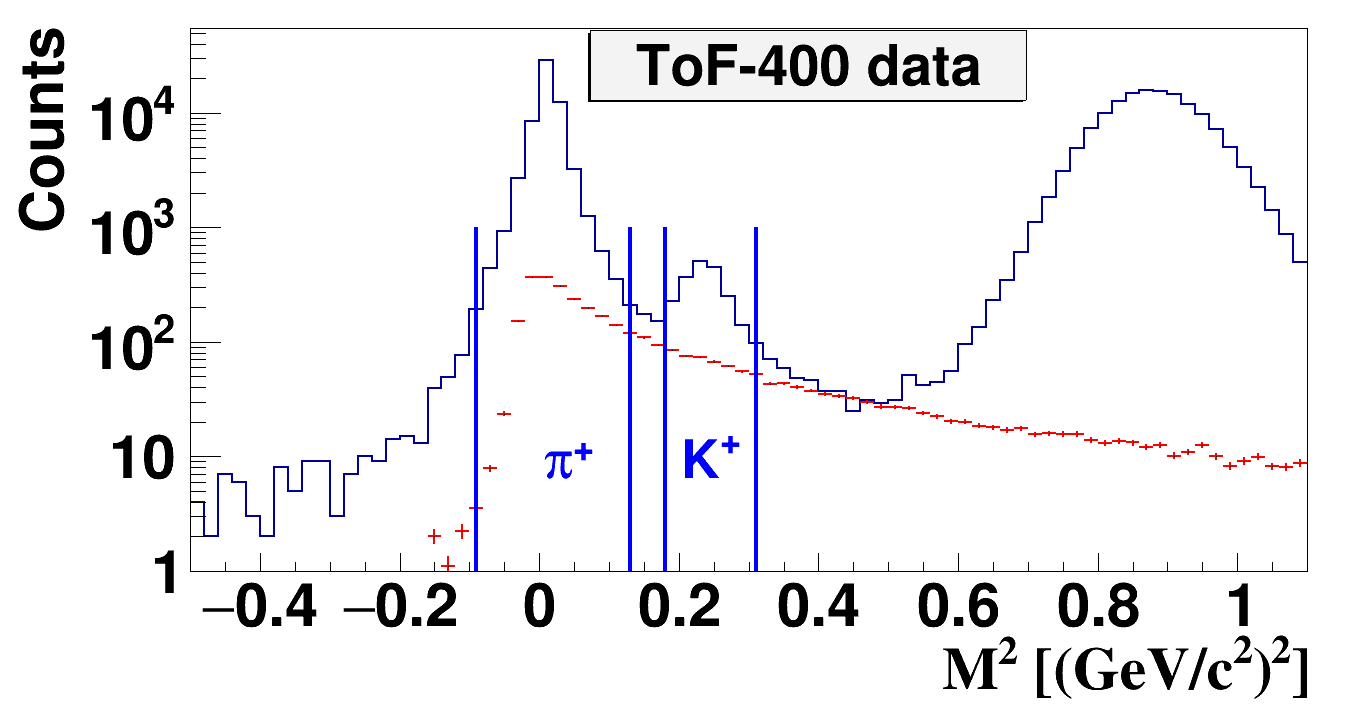}
\includegraphics[width=0.49\textwidth]{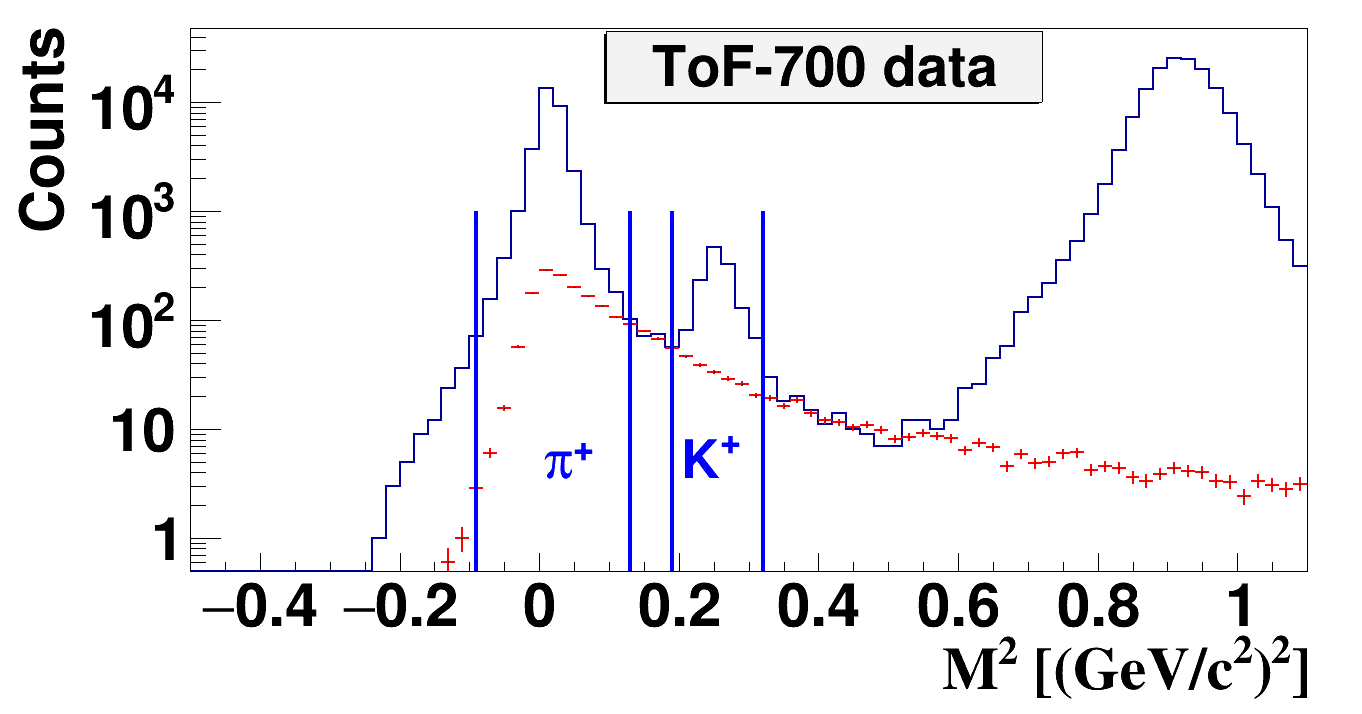}
 \end{center}
 \vspace{-0.7cm}
 \caption{$M^2$ spectra of positively charged particles produced in argon-nucleus
 interactions and measured in the ToF-400 (a) and ToF-700 (b) detectors.
 The vertical lines  show the  ranges of selected $\pi^+$ and $K^+$ mesons.
 The red points show the background estimated from ``mixed events''.}
 \label{m2tof400and700}
\end{figure}

The spectra of the mass squared ($M^2$) of positively charged particles produced
in interactions of the 3.2 AGeV argon beam with various targets are shown in
Figs.~\ref{m2tof400and700}a and \ref{m2tof400and700}b for ToF-400 and ToF-700 data,
respectively. The $\pi^+$ and $K^+$ signals are extracted in the $M^2$ windows
from -0.09 to 0.13 (GeV/c$^2)^2$ and from 0.18 to 0.32 (GeV/c$^2)^2$, respectively.
The signals of $\pi^+$ and $K^+$ and their statistical errors are calculated
according to the formulae: $sig=hist-bg$, $err_{stat}=\sqrt{hist+bg}$, assuming the
background uncertainty is $\sqrt{bg}$. Here $hist$ and $bg$ denote the histogram and
background integral yields within the selected $M^2$ windows.

The shape of the background under the $\pi^+$ and $K^+$ signals in the $M^2$ spectra is
estimated using the ``mixed event'' method. For that, tracks reconstructed in the central
tracker are matched to hits in the ToF detectors taken from different events. The ``mixed event''
background is normalized to the integral of the signal histogram outside the $M^2$ windows
of $\pi^+$ and $K^+$ mesons, i.e in the $M^2$ ranges 0.13-0.18 (GeV/c$^2)^2$ and 0.32-0.4 (GeV/c$^2)^2$.
It was found that the background level differs for light and heavy targets and for different
intervals of rapidity  and transverse momentum.

\begin{figure}[htb]
\begin{center}
\hspace{0.5cm} (a) \hspace{5.5cm} (b)
\includegraphics[width=0.48\textwidth,bb=0 0 751 501]{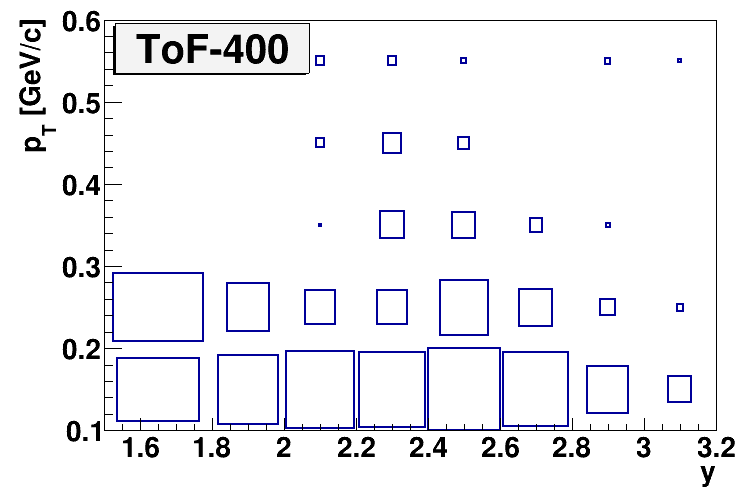}
\includegraphics[width=0.48\textwidth,bb=0 0 749 501]{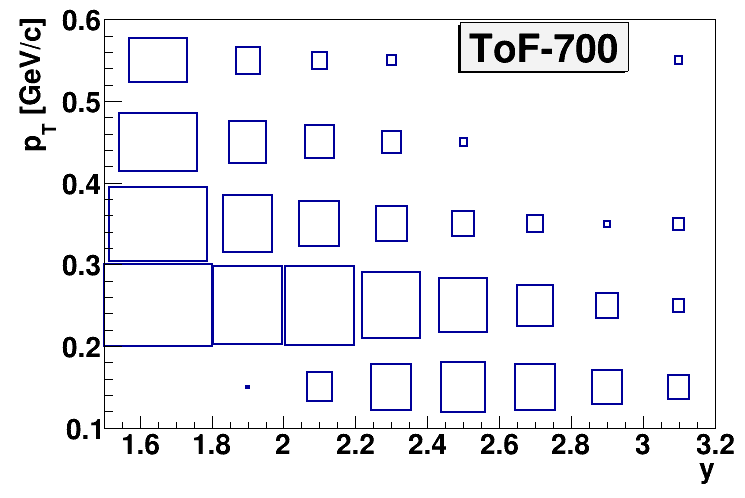}
 \end{center}
 \vspace{-1.0cm}
\caption{Distribution of the $\pi^+$ signals measured in ToF-400 (a)
          and ToF-700 (b) in the rapidity and transverse momentum
           bins in Ar+Sn interactions.}
 \label{ypt_pip_kp_sn}
\end{figure}

The ToF-400  and ToF-700  detectors cover different ranges of rapidity and transverse
momentum of detected particles. Fig.~\ref{ypt_pip_kp_sn} shows the signals of
$\pi^+$ mesons measured in ToF-400 and ToF-700 in the rapidity vs transverse
momentum plane in Ar+Sn interactions before making corrections for the efficiency.

\section{Reconstruction efficiency  and trigger performance}
\label{sect4}

To evaluate the $\pi^+$ and $K^+$ reconstruction efficiency, Monte Carlo data
samples of argon-nucleus collisions were produced with the DCM-SMM event
 generator~\cite{DCM_QGSM,DCM_SMM}. Propagation of particles through the entire
 detector volume and responses of the detectors were simulated using the
 GEANT3 program~\cite{GEANT3} integrated into the BmnRoot software
 framework~\cite{BmnRoot}. To properly describe the GEM detector response
 in the magnetic field, the Garfield++ toolkit \cite{Garfield} for simulation
 of the micropattern gaseous detectors was used.

\begin{figure}[tbh]
\begin{center}
\hspace{1cm} (a) \hspace{7cm} (b)
\includegraphics[width=0.49\textwidth,bb=0 0 536 321]{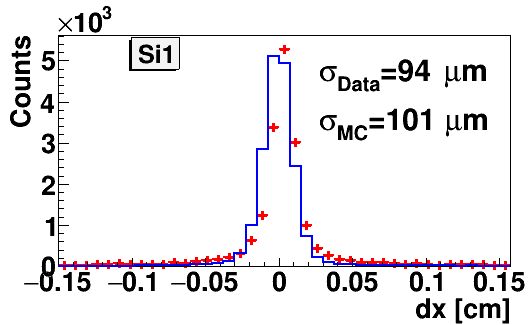}
\includegraphics[width=0.49\textwidth,bb=0 0 536 321]{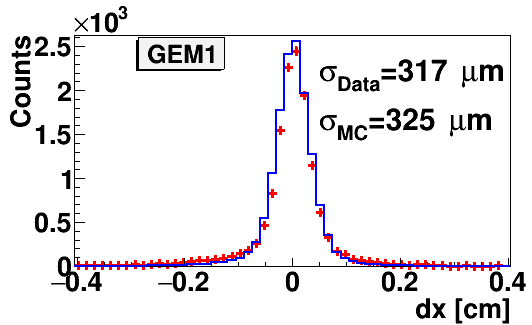}
 \end{center}
 \vspace{-0.6cm}
 \caption {Residual distributions of hits in the X projection (magnet deflection plane)
 with respect to reconstructed tracks: (a)\,-\,in the first forward silicon plane, (b)\,-\,in
 the first GEM plane. The experimental data are shown as red crosses, and the simulated
 data are shown as blue histograms.}
 \label{sigem_resid}
\end{figure}

The efficiencies of the forward silicon, GEM, CSC, DCH and ToF detectors were
adjusted during simulation in accordance with  the measured detector efficiencies.
The Monte Carlo events went through the same chain of reconstruction and identification
as the experimental events.

\begin{figure}[tbh]
 \begin{center}
\vspace{-1.7cm}
\hspace{0cm} (a) \hspace{6cm} (b)

\hspace{-0.5cm}\includegraphics[width=0.45\textwidth,bb=0 0 534 326]{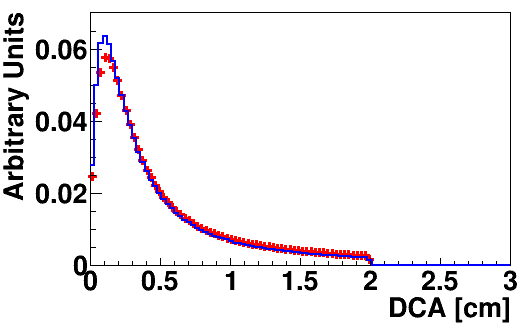}
\includegraphics[width=0.45\textwidth,bb=0 0 534 326]{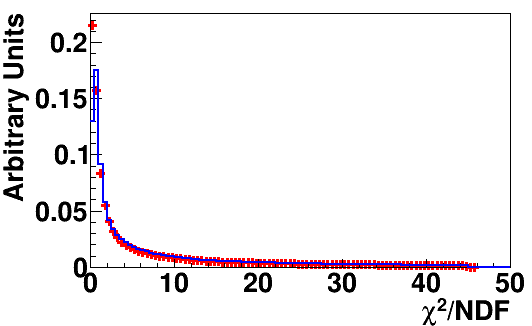}

\hspace{0cm} (c) \hspace{6cm} (d)

\includegraphics[width=0.45\textwidth,bb=0 0 534 326]{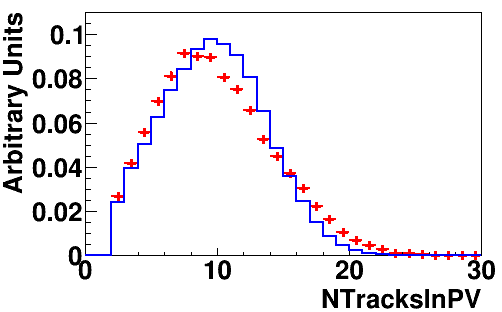}
\includegraphics[width=0.45\textwidth,bb=0 0 534 326]{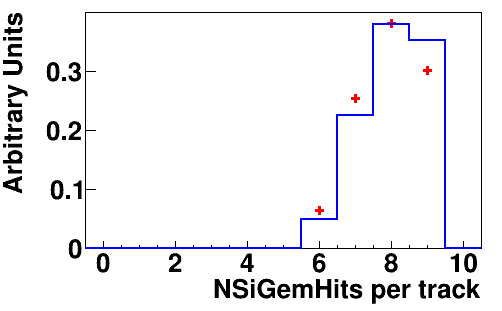}
\end{center}
\vspace{-0.6cm}
 \caption{{Comparison of experimental distributions (red crosses) and GEANT
 distributions of events generated with the DCM-SMM model (blue lines),
 in Ar+A collisions at 3.2 AGeV:
 (a) DCA, distance of closest approach to the primary vertex;
 (b) $\chi^2$/NDF of reconstructed tracks;
 (c)  number of reconstructed tracks in the primary vertex in Ar+Cu interactions;
 (d) Hits per track in the 3 forward Si and 6 GEM detectors.}}
 \label{control_mc_data}
\end{figure}

The level of agreement between the Monte Carlo and experimental distributions is
demonstrated on a set of observables: primary vertices distribution along the
Z-axis (Fig.~\ref{zpv_c}), residuals in the central tracker detectors (Fig.~\ref{sigem_resid}),
closest distance from a track to the primary vertex in the X-Y plane
(DCA), $\chi^2$/NDF, number of reconstructed tracks at the primary vertex and
number of hits per track (Figs.~\ref{control_mc_data}a–d).

\begin{figure}[!htb]
\begin{center}
\hspace{0cm} (a) \hspace{5.0cm} (b)

\hspace{-0.5cm}\includegraphics[width=0.40\textwidth,bb=0 0 646 507]{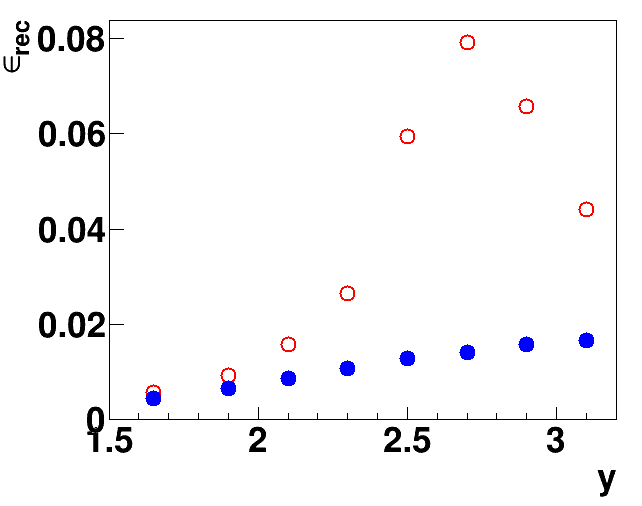}
\includegraphics[width=0.40\textwidth,bb=0 0 648 507]{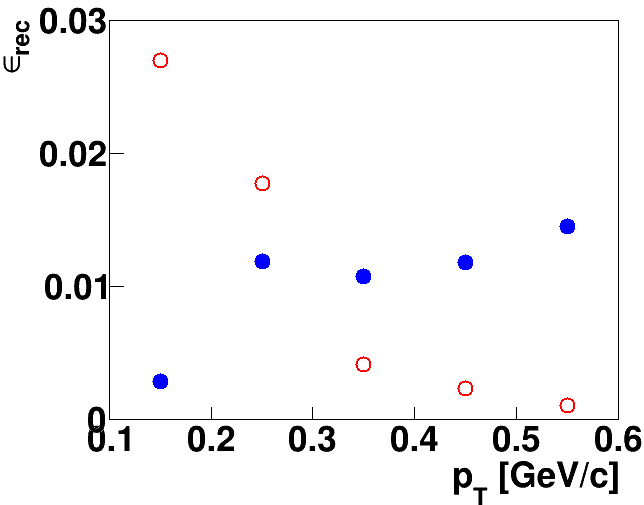}
 \end{center}
 \vspace{-0.6cm}
 \caption{Reconstruction efficiency of $\pi^+$ produced in Ar+Sn collisions,
 detected in ToF-400 (open red circles) and ToF-700 (full blue circles)
 as function of  rapidity $y$  (a) and  $p_T$ (b). The efficiency includes both acceptance and reconstruction.}
 \label{eff}
\end{figure}

The $\pi^+$ and $K^+$ reconstruction efficiencies are calculated in
intervals of rapidity $y$ and transverse momentum $p_T$. The reconstruction
efficiency includes the geometrical acceptance, the detector efficiency,
the kinematic and spatial cuts, the loss of $\pi^+$ and $K^+$ due to in-flight decays and the meson reconstruction.
The reconstruction efficiencies of $\pi^+$ detected in ToF-400 and ToF-700 are
shown in Fig.~\ref{eff} as function of $y$ (left panel) and $p_T$ (right panel) for Ar+Sn interactions.

The trigger efficiency $\epsilon_{trig}$ depends on the number of fired channels
in the BD (SiMD) detectors. It was calculated for events with reconstructed $\pi^+$ and $K^+$
mesons using event samples recorded with an independent trigger based on the SiMD (BD) detectors.
The BD and SiMD detectors cover different and non-overlapping regions of the BM@N acceptance,
that is, they detect different  collision products.
For the BD trigger efficiency estimation, the following relation is used:
$\epsilon_{trig}$ (BD $\ge$ m) = N(BD $\ge$ m $\land$ SiMD $\ge$ n)/N(SiMD $\ge$ n),
where m and n are  the minimum number of fired channels in BD (m = 3, 4) and SiMD (n = 3, 4)
(see Section \ref{sect2}). A similar relation is used to evaluate the SiMD trigger efficiency.
The BD (SiMD) trigger efficiency is averaged over all data with the different values of the
minimum number of fired channels in SiMD (BD).

\begin{figure}[tbh]
\begin{center}
\vspace{0.0cm}
\hspace{-1cm} \includegraphics[width=0.45\textwidth,bb=0 0 644 505]{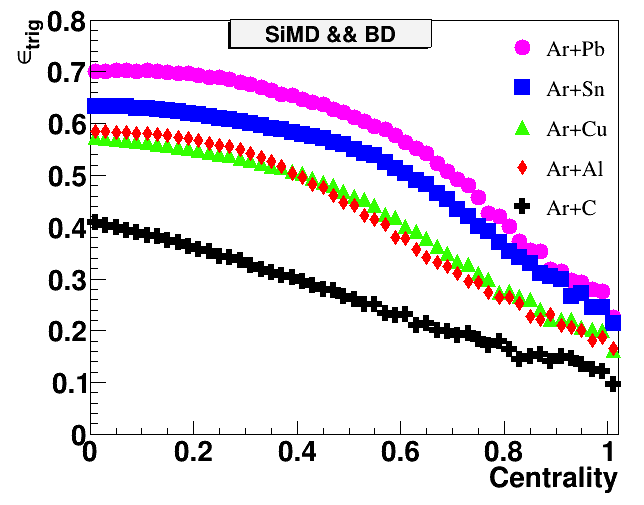}
\end{center}
\vspace{-0.9cm}
\caption {Trigger efficiency for interactions of the argon beam with various
targets (C, Al, Cu, Sn, Pb) with a reconstructed $\pi^+$ as a function of the
event centrality estimated from the simulation.}
\label{TrigEff}
\end{figure}

The efficiency of the combined BD and SiMD triggers was calculated as the product of
the efficiencies of the BD and SiMD triggers.
The trigger efficiency, for events with a reconstructed $\pi^+$, averaged over
all data collected with the trigger conditions 1) BT$\land$(BD $\ge\rm{3,4}$);
2) BT$\land$(SiMD $\ge\rm{3,4}$);
3) BT$\land$(BD $\ge\rm{2}$)$\land$(SiMD $\ge\rm{3}$) (see Section \ref{sect2})
is shown in Fig.~\ref{TrigEff} as a function of the event centrality estimated from
simulation. The event  centrality is determined as the fraction of
the interaction cross section in the interval $\mathrm{[0,b]}$  of the impact
parameter $\mathrm b$ of the nucleus-nucleus collision to the total interaction
cross section. It is clearly seen that the trigger efficiency decreases with a decrease
in the mass of the target and an increase in the centrality of the collision.
The trigger efficiency for events with a reconstructed $K^+$ was found to be slightly
higher, 6\% higher in Ar+C collisions and 11\% higher for Ar+Pb collisions for the
combined trigger BT$\land$(BD $\ge\rm{2}$)$\land$(SiMD $\ge\rm{3}$).

\section{Cross sections, multiplicities, and systematic uncertainties}
\label{sect5}

The $\pi^+$ ($K^+$) mesons in Ar+C, Al, Cu, Sn, Pb interactions are measured in the
following kinematic range: transverse momentum  $0.1<p_T<0.6$~GeV/c ($0.1<p_T<0.5$~GeV/c)
and rapidity in the laboratory frame $1.5<y<3.2$ ($1.0 <y<2.0$). The analysis takes into
account the  track dependence of the trigger efficiency. No significant variation in the
reconstruction efficiency with the track multiplicity was found.
The differential cross sections {\small $d^2\sigma_{\pi,K}(y,p_T)/dydp_T$} and
multiplicities {\small $d^2N_{\pi,K}(y,p_T)/dydp_T$ of $\pi^+$} and $K^+$ meson
production in Ar+C, Al, Cu, Sn, Pb interactions are calculated using the relations:

\begin{small}
\vspace{0cm}
$d^2\sigma_{\pi,K}(y,p_T)/dydp_T =  \Sigma [ d^2 n_{\pi,K}(y,p_T, N_{tr}) / (\epsilon_{trig}(N_{tr}) dy dp_T)] \times 1 / ( L \epsilon_{rec}(y,p_T))$

$d^2N_{\pi,K}(y,p_T)/dydp_T = d^2\sigma_{\pi,K}(y,p_T) / (\sigma_{inel} dydp_T)$
\hspace{4cm} (1)
 \end{small}
\noindent where the sum is performed over bins of the number of tracks in the
primary vertex, $N_{tr}$, $n_{\pi,K}(y, p_T, N_{tr})$ is the number of
reconstructed $\pi^+$ or $K^+$ mesons in the  intervals $dy$ and $dp_T$,
$\epsilon_{trig}(N_{tr}) $ is the track-dependent trigger efficiency,
$\epsilon_{rec}$ is the reconstruction efficiency of $\pi^+$ or $K^+$, $L$
is the luminosity,   and $\sigma_{inel}$ is the inelastic cross section for argon-nucleus interactions.

Table~\ref{uncertainties_tablepiK} summarizes the mean values, averaged over $p_T$, $y$, and $N_{tr}$,
of the systematic uncertainties of the various factors of Eq. (1), $n_{\pi,K}$, $\epsilon_{rec}$,
and $\epsilon_{trig}$. Details are given below, including the uncertainty of the luminosity measurement.
The model uncertainty of $\sigma_{inel}$ is given in Table \ref{extrap_tablepiK}.

Several sources are considered for the evaluation of the systematic uncertainty
of the $\pi^+$ and $K^+$ yield, $n_{\pi,K}$, and the reconstruction
efficiency $\epsilon_{rec}$. The most significant ones are discussed below.
Some of them affect both the yield $n_{\pi,K}$ and the reconstruction
efficiency, $\epsilon_{rec}$. For these cases the correlated effect is taken
into account by the variations on the $n_{\pi,K}/\epsilon_{rec}$ ratio:
\begin{itemize}
\vspace{-0.3cm}
\item Systematic uncertainty of the central tracking detector efficiency: it is
estimated from the remaining difference in the number of track hits in the central
 detectors in the simulation relative to the data (see Fig. \ref{control_mc_data}d) and found to be within 3\%.
\vspace{-0.3cm}
\item Systematic uncertainty of the matching of central tracks to the CSC (DCH)
hits and ToF-400 (ToF-700) hits: it is estimated from the remaining difference
in the matching efficiency in the simulation relative to the data and found to be within 5\%.
\vspace{-0.3cm}
\item Systematic uncertainty of the reconstruction efficiency due to the remaining difference
      in the X/Y distribution of primary vertices in the simulation relative to the data.
\vspace{-0.3cm}
\item Systematic uncertainty of the background subtraction in the mass-squared $M^2$
spectra of identified particles: it is estimated as the difference between the
background integral under the meson windows taken from ``mixed events''
(as described in Section~\ref{sect3}) and from the fitting of the $M^2$ spectra by a
linear function. The latter is done in the $M^2$ range -0.14-0.4 (GeV/c$^2)^2$,
excluding the $\pi^+$ and $K^+$ windows.
\vspace{-0.3cm}
\end{itemize}
\noindent The total systematic uncertainty of the  yield and reconstruction
efficiency for the various targets, calculated as the quadratic sum of these
uncertainties, is listed in Table~\ref{uncertainties_tablepiK}.

The luminosity is calculated from the beam flux $\Phi$ as given by the beam trigger
(see Section \ref{sect2}) and the target thickness $l$ using the relation: $ L = \Phi \rho l$ where $\rho $
is the target density expressed in atoms/cm$^3$. The systematic uncertainty of the luminosity
is estimated from the fraction of the beam which can miss the target, determined from the
vertex positions, and found to be within 2\%.

\begin{table}[!hbp]
  \caption{
           {Mean systematic uncertainties in $y$, $p_T$ bins of the $\pi^+$ and $K^+$
	   mesons measured in argon-nucleus interactions (see text for details).}}
\begin{footnotesize}
\vspace{0.3cm}
\hspace{2.5cm}
\begin{tabular}{|l|c|c|c|c|c|}
\hline
\multicolumn{1}{|r|}{} & Ar+C & Ar+Al & Ar+Cu & Ar+Sn & Ar+Pb \\
                                  &  \%    & \%      & \%       &  \%      &    \% \\
\hline
 $\pi^+$                       &      &     &     &      &    \\
$n_{\pi}$,  $\epsilon_{rec}$   &  14  & 12  & 12  & 10   & 10 \\
$\epsilon_{trig}$              &   9  &  7  &  7  &  7   &  7 \\
 Total                         &   17 &  14 &  14 &  13  &  13 \\
 & & & & & \\
$K^+$                          &      &     &     &      &   \\
 $n_{K}$,  $\epsilon_{rec}$    &   25 &  23 &  14 &  13  &  15 \\
 $\epsilon_{trig}$             &   31 &  14 &   9 &   8 &  8  \\
Total                          &   40 &  27 &  17 &  16 & 17 \\
\hline
\end{tabular}
\end{footnotesize}
\label{uncertainties_tablepiK}
\end{table}

For the evaluation of the systematic uncertainty of the trigger efficiency $\epsilon_{trig}$,
the following sources are considered:
\begin{itemize}
\vspace{-0.3cm}
\item The systematic uncertainty associated with the factorization assumption of
the two trigger factors, BD and SiMD,  was estimated from the difference
of $\epsilon_{trig}$ evaluated as described in Section \ref{sect4}, with
the result evaluated using the limited amount of events registered with the beam trigger BT.
\vspace{-0.3cm}
\item To estimate a possible distortion of $\epsilon_{trig}$ (BD $\ge$ m)  due to
the selection of events with the hardware-set condition N(SiMD $\ge$ n), $\epsilon_{trig}$
was also evaluated using the events recorded with the beam trigger BT. The difference
between the results is treated as another source of systematic uncertainty of the trigger efficiency.
\vspace{-0.3cm}
\item Variations of the trigger efficiency on the track multiplicity in the primary
vertex and on the X/Y vertex position.
 \end{itemize}
\vspace{-0.3cm}
\noindent The total systematic uncertainty of the trigger efficiency for the various
targets, calculated as the quadratic sum of these uncertainties, is listed in Table~\ref{uncertainties_tablepiK}.

The inelastic cross sections of Ar+C, Al, Cu, Sn, Pb interactions are taken from the predictions
of the DCM-SMM model which are consistent with the results calculated by the formula:
 $\sigma_{inel} = \pi R_0^2 (A_P^{1/3} + A_T^{1/3})^2$, where $R_0 = 1.2$~fm is the
 effective nucleon radius, $A_P$ and $A_T$ are the atomic numbers of the projectile
 and target nucleus \cite{HadesL0}.
 The systematic uncertainties for the Ar+C, Al, Cu, Sn, Pb  inelastic cross sections are estimated
 from an alternative formula \cite{AngelovCC} which approximates the measured nucleus-nucleus cross sections:
  $\sigma_{inel} = \pi R_0^2 (A_P^{1/3} + A_T^{1/3} - b)^2$ with $R_0 = 1.46$ fm and $b = 1.21$
The values and uncertainties of $\sigma_{inel}$ for Ar+C, Al, Cu, Sn, Pb interactions
are given in Table~\ref{extrap_tablepiK}.

\section{Results and discussion}
\label{sect6}

\begin{figure}[tbh]
\begin{center}
\vspace{-2.0cm}
\includegraphics[width=0.45\textwidth,bb=0 0 600 600]{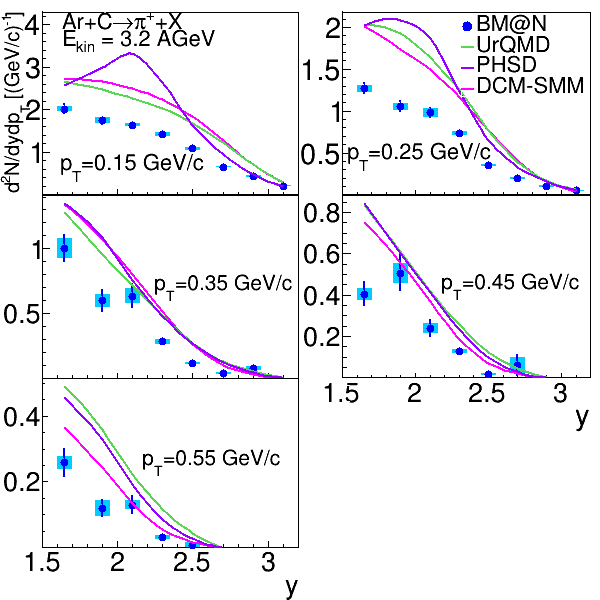}
\includegraphics[width=0.45\textwidth,bb=0 0 600 600]{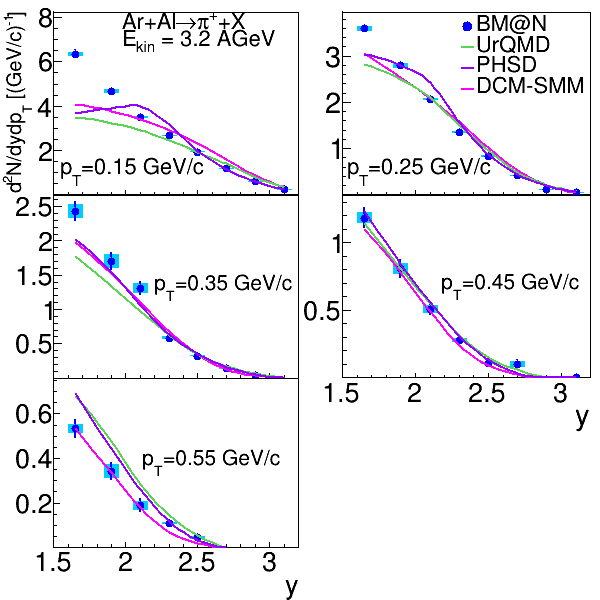}
\includegraphics[width=0.45\textwidth,bb=0 0 600 600]{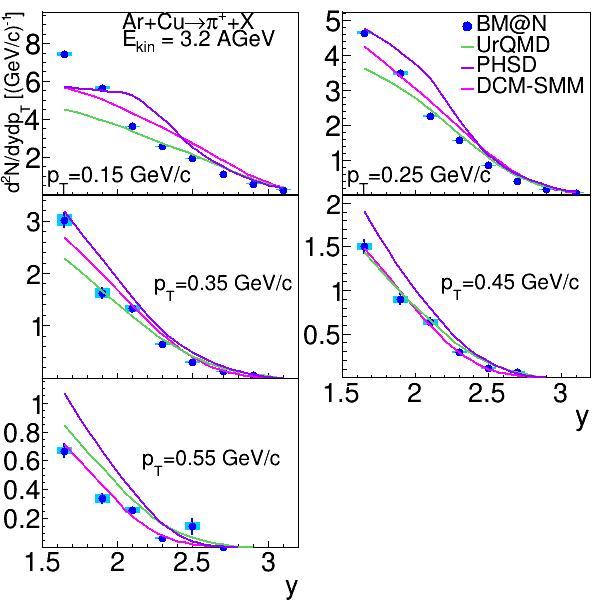}
\includegraphics[width=0.45\textwidth,bb=0 0 600 600]{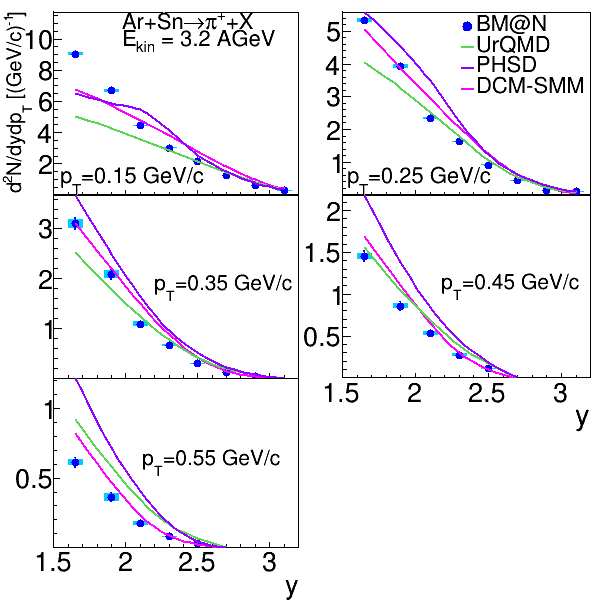}
\includegraphics[width=0.45\textwidth,bb=0 0 600 600]{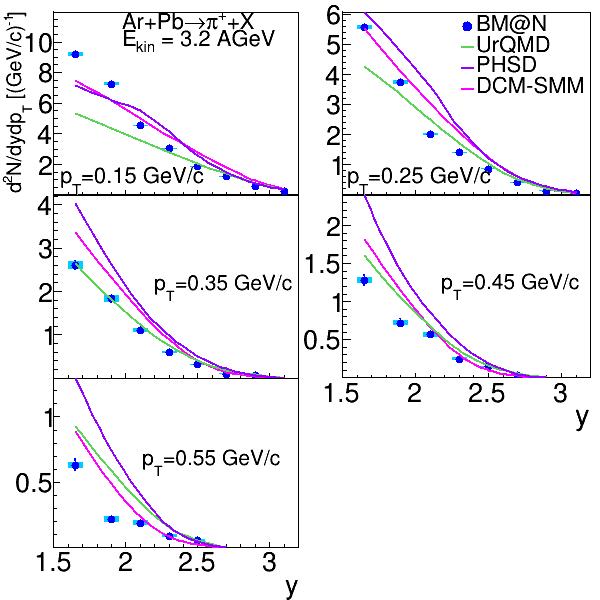}
\end{center}
\vspace{-0.8cm}
 \caption{Rapidity spectra ($y$) of $\pi^+$ mesons produced in  Ar+C, Al, Cu, Sn, Pb
 interactions at a kinetic energy of 3.2~AGeV. The results are presented for
 different $p_T$ bins.  The vertical bars and boxes represent the statistical
 and systematic uncertainties, respectively. The predictions of the DCM-SMM,
 UrQMD and PHSD models are shown as rose, green and magenta lines.}
 \label{yields_ypi}
\end{figure}

\begin{figure}[tbh]
\begin{center}
\vspace{-3.0cm}
\includegraphics[width=0.36\textwidth,bb=0 0 300 600]{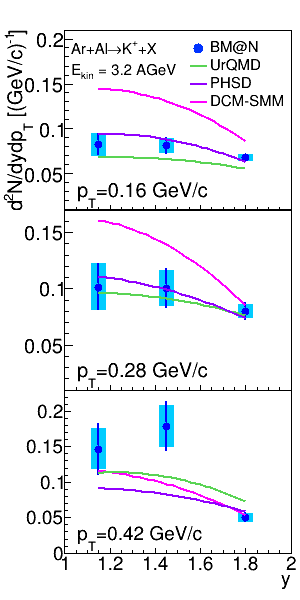}
\includegraphics[width=0.36\textwidth,bb=0 0 300 600]{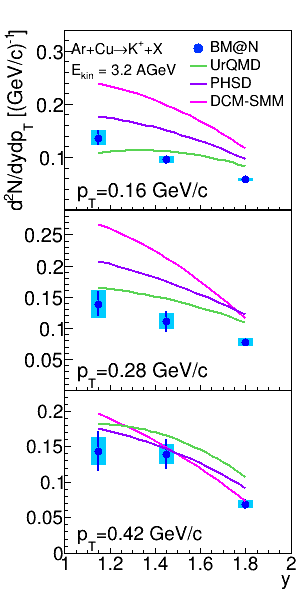}
\includegraphics[width=0.36\textwidth,bb=0 0 300 600]{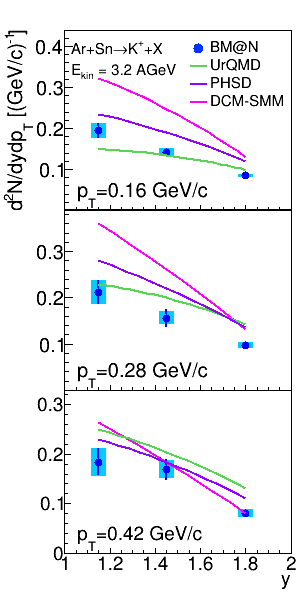}
\includegraphics[width=0.36\textwidth,bb=0 0 300 600]{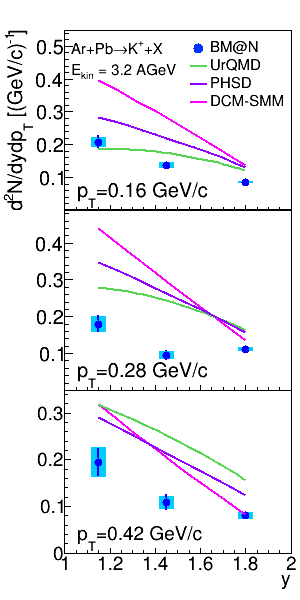}
\end{center}
\vspace{-0.8cm}
 \caption {Rapidity spectra ($y$) of $K^+$ mesons produced in  Ar+Al, Cu, Sn, Pb interactions
 at a kinetic energy of 3.2~AGeV. The results are presented for different $p_T$ bins.
 The vertical bars and boxes represent the statistical and systematic uncertainties,
 respectively. The predictions of the DCM-SMM, UrQMD and PHSD models are shown as rose,
  green and magenta lines.}
 \label{yields_yK}
\end{figure}

The rapidity spectra of $\pi^+$ and $K^+$ mesons are shown in Figs.~\ref{yields_ypi} and~\ref{yields_yK},
 respectively, for different $p_T$ bins for various targets. At a kinetic energy of 3.2 GeV/nucleon,
 the rapidity of the nucleon-nucleon center-of-mass (CM) system is $y_{CM}=1.08$. The rapidity
 intervals covered in the present measurements, $y = 1.5-3.2$ and $y = 1.0-2.0$ for  $\pi^+$ and $K^+$,
  respectively, correspond therefore to the forward and central rapidity regions in the nucleon-nucleon
 CM system.
Figs.~\ref{yields_ypi} and~\ref{yields_yK} show also a comparison of the experimental results
with the predictions of the DCM-SMM~\cite{DCM_QGSM,DCM_SMM}, UrQMD~\cite{UrQMD} and PHSD~\cite{PHSD} models.
For $\pi^+$, the three models have quite similar predictions, in particular at medium and
high $p_T$ bins for all targets, except for the C target (and to a lesser extent also
the Al target) where the PHSD model is markedly different from the DCM-SMM and UrQMD models
at mid-rapidity. The three models are in reasonable agreement with the experimental results
at forward rapidity and high $p_T$ for all targets. In general, for the heavier targets, the
models overshoot the data at mid-rapidity and high-$p_T$, the UrQMD predictions being closer
to the data. All three models fail to reproduce the shape and magnitude of the C data at mid-rapidity.
For $K^+$, there are large differences, up to a factor of $\sim$2, between the three models.
In general, the PHSD and DCM-SMM models over-predict the data for all $p_T$ bins, whereas the
UrQMD is closer to the data.

\begin{figure}[tbh]
\begin{center}
\vspace{-1.2cm}
\includegraphics[width=0.48\textwidth,bb=0 0 720 520]{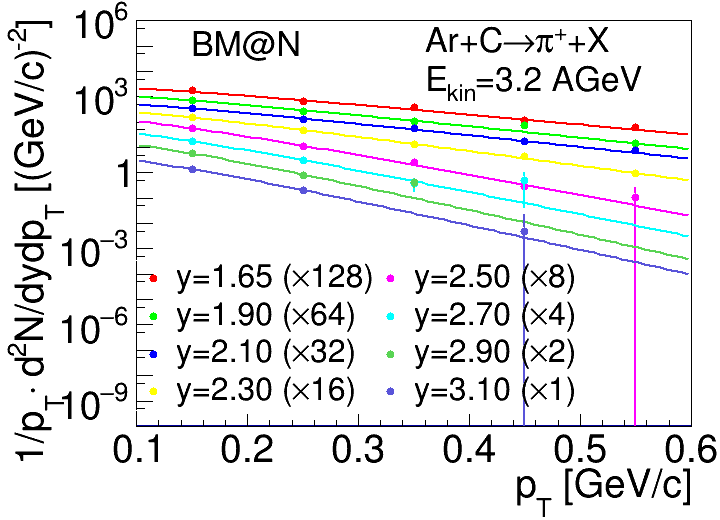}
\includegraphics[width=0.48\textwidth,bb=0 0 720 520]{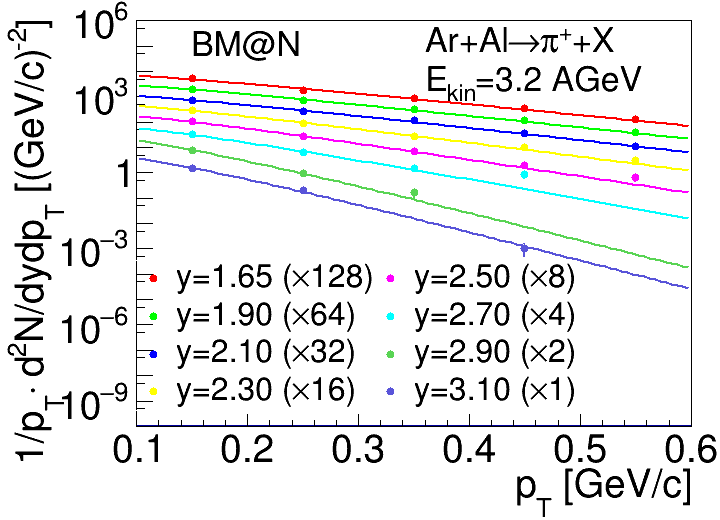}
\includegraphics[width=0.48\textwidth,bb=0 0 720 520]{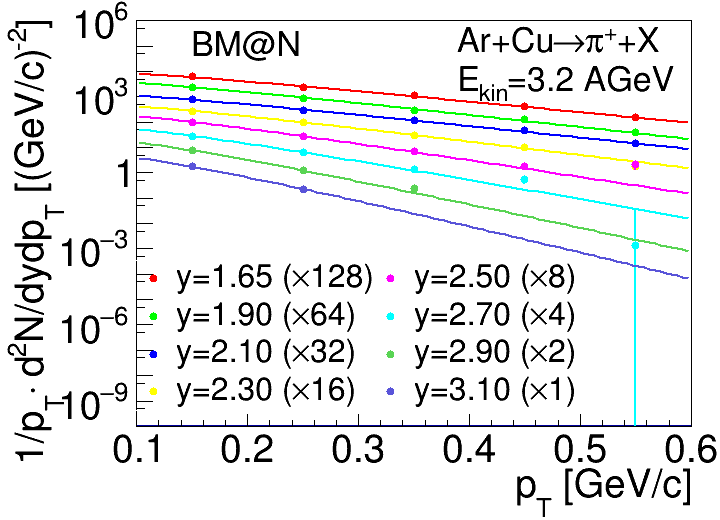}
\includegraphics[width=0.48\textwidth,bb=0 0 720 520]{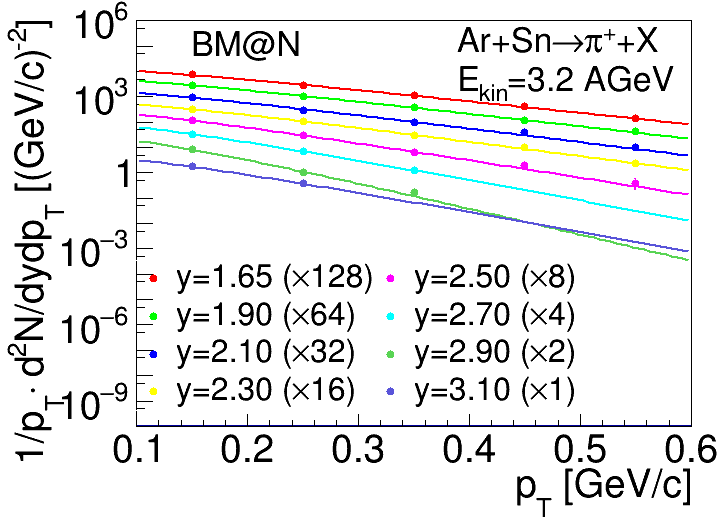}
\includegraphics[width=0.48\textwidth,bb=0 0 720 520]{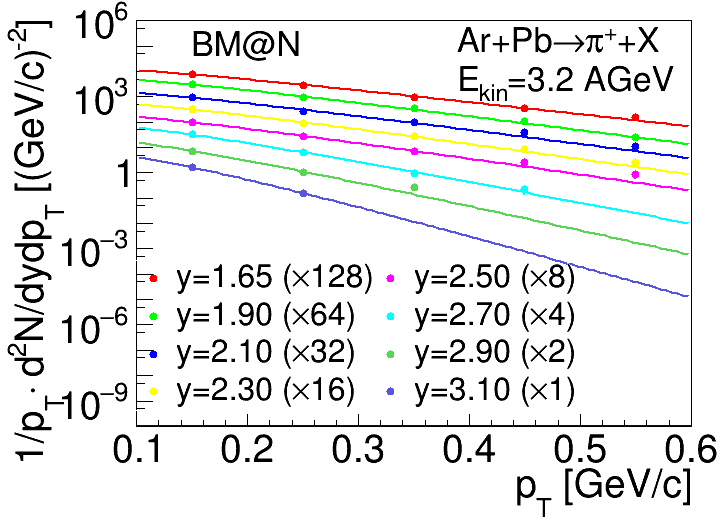}
\end{center}
\vspace{-0.8cm}
 \caption{Transverse momentum ($p_T$) spectra of $\pi^+$ mesons produced in Ar+C, Al, Cu, Sn, Pb
 interactions at 3.2~AGeV.
 The results are given for bins of the $\pi^+$ rapidity. The lines
 represent the results of the parametrization described in the text.}
 \label{yields_ptpi}
\end{figure}

\begin{figure}[tbh]
\begin{center}
\vspace{-3.0cm}
\includegraphics[width=0.36\textwidth,bb=0 0 299 586]{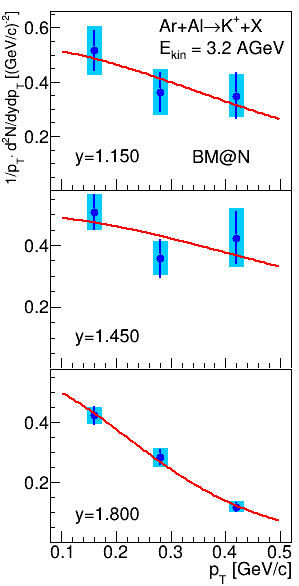}
\includegraphics[width=0.36\textwidth,bb=0 0 299 586]{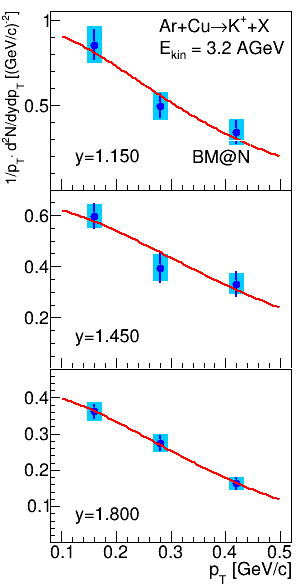}
\includegraphics[width=0.36\textwidth,bb=0 0 299 586]{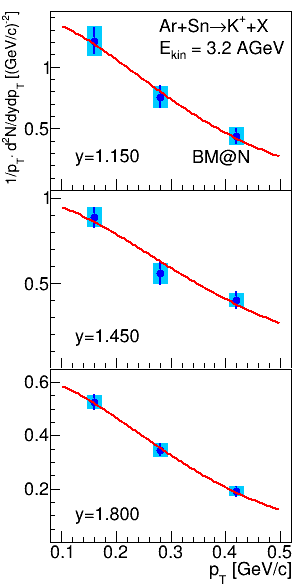}
\includegraphics[width=0.36\textwidth,bb=0 0 299 586]{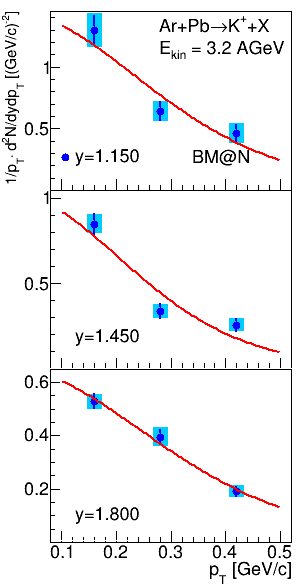}
\end{center}
\vspace{-0.8cm}
\caption{Transverse momentum ($p_T$) spectra of $K^+$ mesons produced in Ar+Al, Cu, Sn, Pb
interactions at 3.2~AGeV.
The results are given for three bins of the $K^+$ rapidity. The vertical bars and boxes
represent the statistical and systematic uncertainties, respectively. The lines represent the
results of the parametrization described in the text.}
\label{yields_ptK}
\end{figure}

The  $p_T$ spectra of $\pi^+$ and $K^+$ mesons are shown in
Figs.~\ref{yields_ptpi} and \ref{yields_ptK}, respectively, for different rapidity
$y$ bins and for all targets.
Due to the low statistics of the $K^+$ yield in Ar+C interactions, the
results are shown only for the entire measured ranges of $y$ and $p_T$.

\begin{figure}[tbh]
\begin{center}
\vspace{-1.0cm}
\includegraphics[width=0.48\textwidth,bb=0 0 718 493]{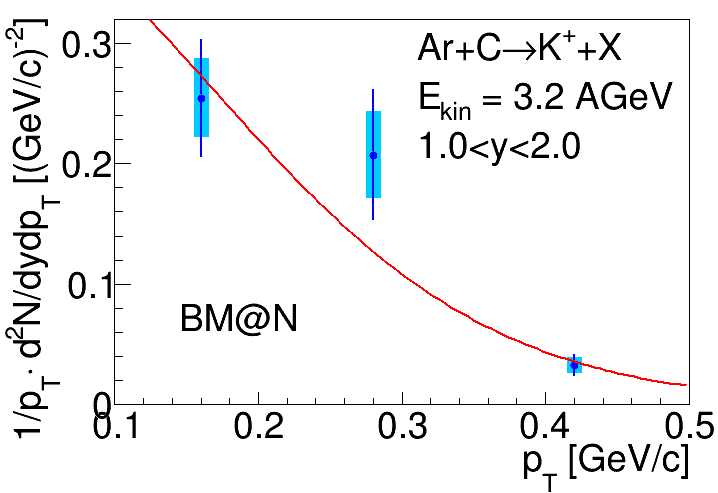}
\includegraphics[width=0.48\textwidth,bb=0 0 718 493]{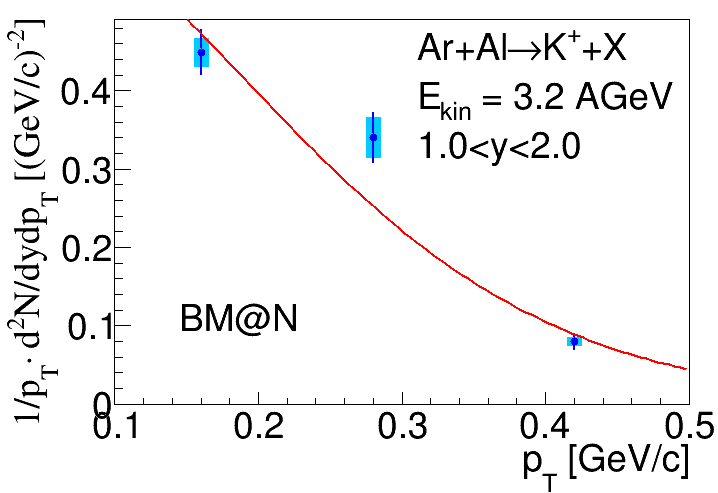}
\includegraphics[width=0.48\textwidth,bb=0 0 718 493]{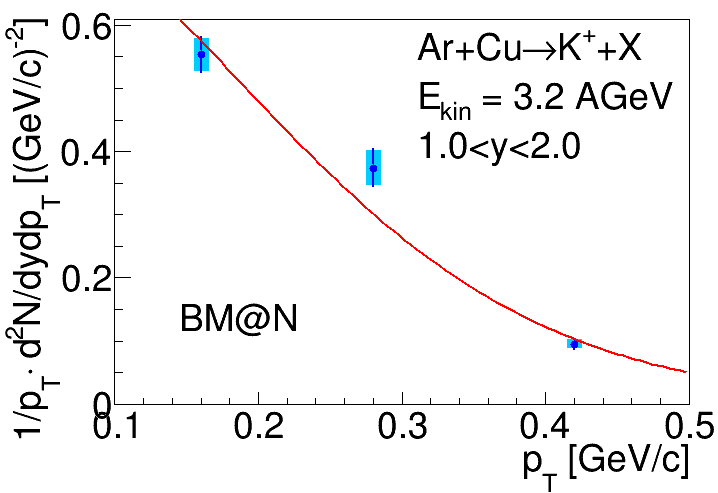}
\includegraphics[width=0.48\textwidth,bb=0 0 718 493]{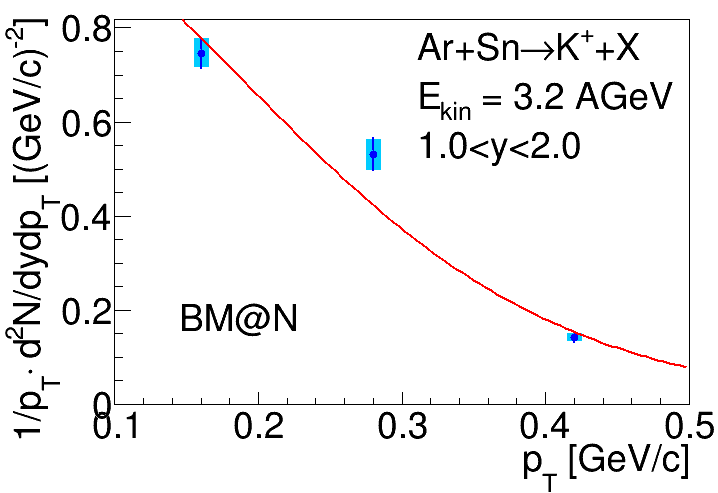}
\includegraphics[width=0.48\textwidth,bb=0 0 718 493]{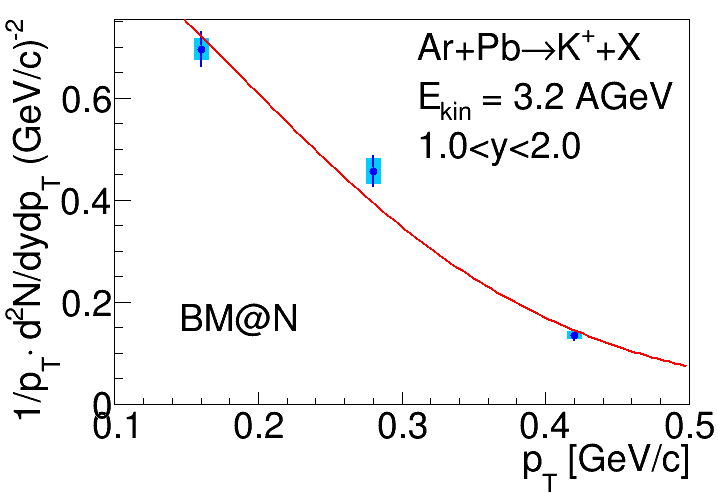}
\end{center}
\vspace{-0.8cm}
\caption{Transverse momentum ($p_T$) spectra of $K^+$ mesons produced in Ar+C, Al, Cu, Sn, Pb
interactions at 3.2~AGeV for the entire measured $K^+$ rapidity range. The vertical bars and
boxes represent the statistical and systematic uncertainties, respectively. The  lines
represent the results of the parametrization described in the text.}
\label{yields_ptKfull}
\end{figure}

\begin{figure}[tbh]
\begin{center}
\vspace{-1.2cm}
\includegraphics[width=0.48\textwidth,bb=0 0 718 493]{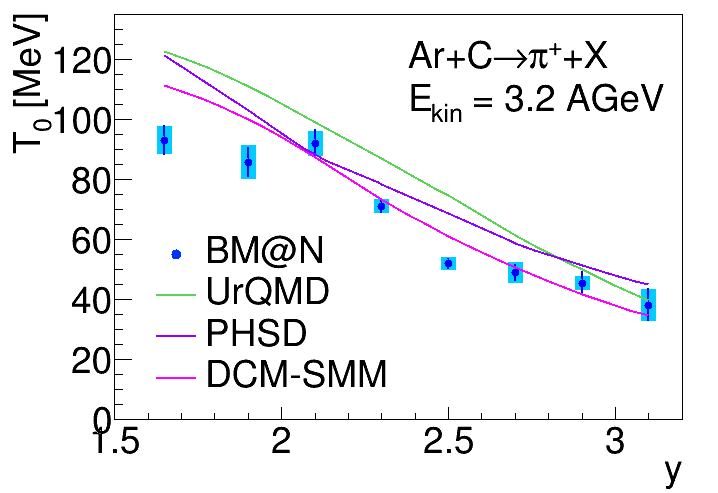}
\includegraphics[width=0.48\textwidth,bb=0 0 718 493]{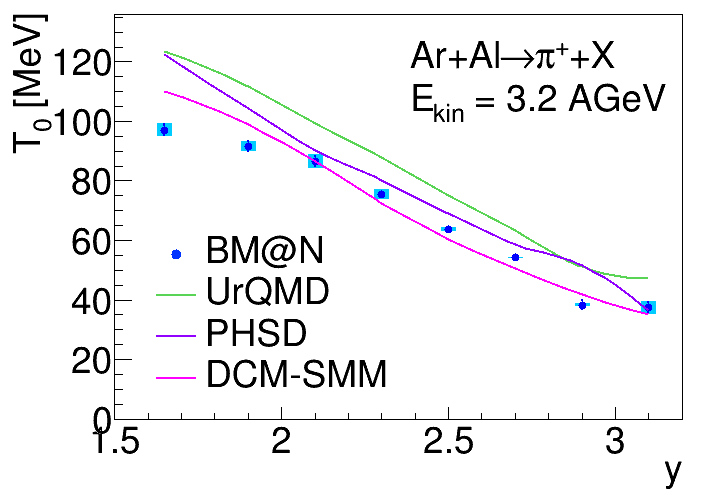}
\includegraphics[width=0.48\textwidth,bb=0 0 718 493]{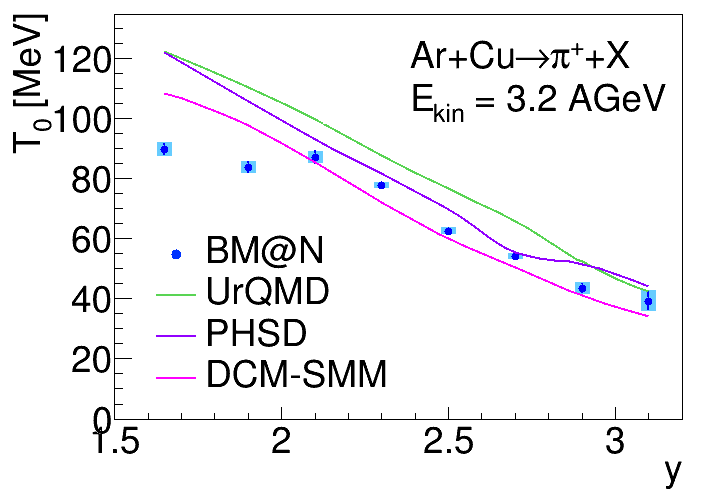}
\includegraphics[width=0.48\textwidth,bb=0 0 718 493]{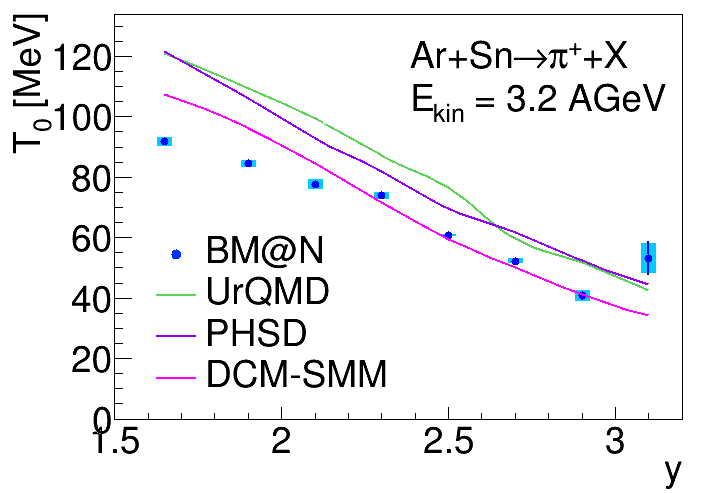}
\includegraphics[width=0.48\textwidth,bb=0 0 718 493]{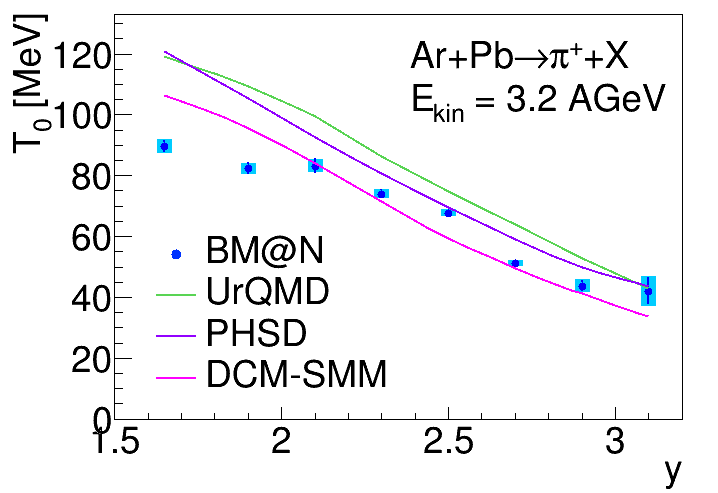}
\end{center}
\vspace{-0.8cm}
 \caption{Rapidity $y$ dependence of the inverse slope parameter $T_0$ determined
 from the fits of the $\pi^+$ $p_T$ spectra in Ar+C, Al, Cu, Sn, Pb interactions.
 The vertical bars and boxes represent the statistical and systematic uncertainties,
 respectively. The predictions of the DCM-SMM, UrQMD and PHSD models are shown as rose, green and magenta lines.}
 \label{T0pi_rapidity}
\end{figure}

The  $p_T$ spectra of $K^+$ mesons integrated over the entire measured
rapidity range are shown in Fig.~\ref{yields_ptKfull}.
In Figs.~\ref{yields_ptpi} $-$ \ref{yields_ptKfull},  the $p_T$ spectra are
parameterised by an exponential function as:

$1/p_T \cdot d^2N/dydp_T\propto {\rm exp}(-(m_T-m_{\pi,K})/T_0)$,

\noindent where $m_T=\sqrt{m_{\pi,K}^2+p_T^2}$ is the transverse mass,
and the inverse slope, $T_0$,  is a fitting parameter. The $T_0$ values
obtained  from the fits of the $\pi^+$ spectra  are shown in Fig.~\ref{T0pi_rapidity}.
The $T_0$ values are about 40~MeV at the most forward rapidity $y \approx 3$,
rising to 90 MeV toward more central rapidities at $y \approx 1.6$.
In general, the $y$ dependence of the fitting results for $\pi^+$ mesons is consistent
with  the predictions of the DCM-SMM, UrQMD and PHSD models, but the experimental
results exhibit a flatter dependence of the $T_0$ values in the
central rapidity range as opposed to the rising dependence of the inverse slopes
predicted by the models.

\begin{figure}[tbh]
\begin{center}
\vspace{-1.2cm}
\includegraphics[width=0.48\textwidth,bb=0 0 718 493]{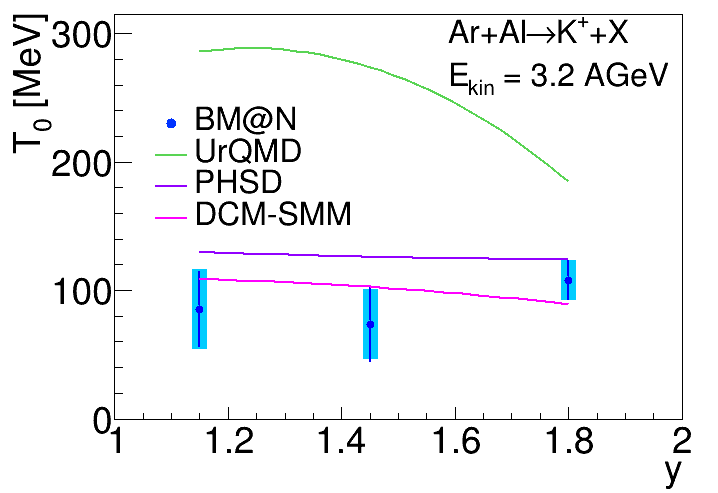}
\includegraphics[width=0.48\textwidth,bb=0 0 718 493]{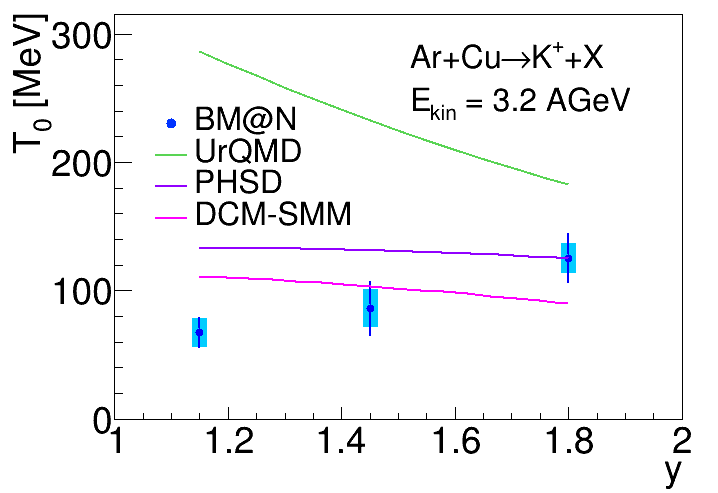}
\includegraphics[width=0.48\textwidth,bb=0 0 718 493]{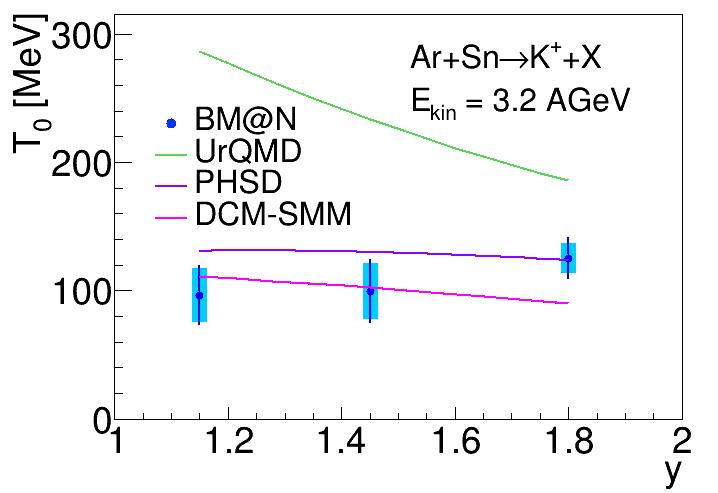}
\includegraphics[width=0.48\textwidth,bb=0 0 718 493]{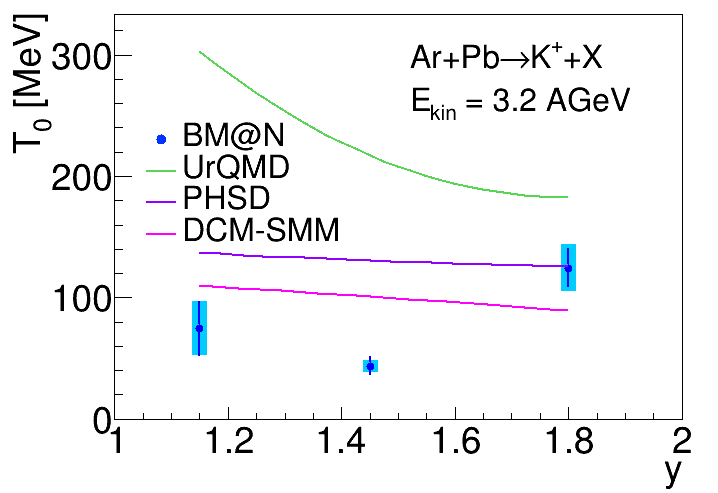}
\end{center}
\vspace{-0.8cm}
 \caption{Rapidity $y$ dependence of the inverse slope parameter $T_0$ extracted from the
 fits of the $K^+$ $p_T$ spectra in Ar+Al, Cu, Sn, Pb interactions. The vertical bars and
  boxes represent the statistical and systematic uncertainties, respectively. The predictions
  of the DCM-SMM, UrQMD and PHSD models are shown as rose, green and magenta lines.}
 \label{T0K_rapidity}
\end{figure}

The $T_0$ values for $K^+$ mesons obtained in 3 $y$ bins  are shown in
Fig. ~\ref{T0K_rapidity}. In spite of the large statistical and systematic
errors,  $T_0$ exhibits a rather weak dependence on $y$.  The $T_0$ values
 for the entire measured range of $1.0 < y < 2.0$ are consistent, within the
 experimental uncertainties, with 80 MeV for all the targets (see Table \ref{yield_tablepiK}).
 The weak dependence of $T_0$ on $y$ is reproduced by the PHSD and DCM-SMM models,
 the latter being in general closer to the data. The UrQMD predicts a strong
 dependence on $y$, with $T_0$ values much larger than the measured ones.

\begin{table}[!hbp]
  \caption{$\pi^+$ and $K^+$ meson multiplicities measured  in Ar+C, Al, Cu, Sn, Pb
  interactions at the argon beam energy of 3.2 AGeV. The first error is statistical,
  the second one is systematic. The third error, given for the full $\pi^+$ and $K^+$
  multiplicities, is the model uncertainty in the extrapolation factor to the
  full phase space (see text and Table \ref{extrap_tablepiK}). }
\begin{footnotesize}
\vspace{0.3cm}
\hspace{-0.5cm}
\begin{tabular}{|l|c|c|c|c|c|}
\hline
& & & & & \\
3.2 AGeV argon & Ar+C & Ar+Al & Ar+Cu & Ar+Sn & Ar+Pb \\
beam & & & & & \\
& & & & & \\
Measured $\pi^+$ & $0.42 \pm 0.008 \pm$ & $ 1.00 \pm 0.01 \pm$  & $1.14 \pm 0.01 \pm$  & $1.28 \pm 0.01 \pm$ & $1.25 \pm 0.01 \pm$ \\
mult. $N_{\pi+}$ & $0.045$ & $0.07$ & $0.08$ & $0.09$ & $0.08$ \\
& & & & & \\
Measured $K^+$ & $1.59 \pm 0.29 \pm$ & $3.90 \pm 0.28 \pm$ & $4.17 \pm 0.21 \pm$ & $5.60 \pm 0.22 \pm$ & $5.10 \pm 0.22 \pm$ \\
mult. $N_{K+}/10^{-2}$ & $0.65$ & $0.61$ & $0.66$ & $0.75$ & $0.92$ \\
& & & & & \\
Full $\pi^+$ mult. & $1.365 \pm 0.026 \pm$ & $3.73 \pm 0.04 \pm$  & $5.07 \pm 0.04 \pm$ & $ 6.55 \pm 0.05 \pm$ & $ 7.39 \pm 0.06 \pm$ \\
$N_{\pi+}^{tot}$ & $0.146 \pm 0.08$ & $0.26 \pm 0.13$ & $0.36 \pm 0.08$ & $0.46 \pm 0.33$ & $0.47 \pm 0.69$ \\
 & & & & & \\
Full $K^+$ mult. & $4.47 \pm 0.81 \pm$ & $11.8 \pm 0.9 \pm$  & $13.9 \pm 0.7 \pm$  & $20.7 \pm 0.8 \pm$  & $ 20.9 \pm 0.9 \pm$ \\
$N_{K+}^{tot}/10^{-2}$ & $1.83 \pm 1.05$ & $1.8 \pm 2.6$ & $2.2 \pm 2.7$ & $2.8 \pm 3.3$ & $3.8 \pm 2.2$ \\
 & & & & & \\
$N_{K+}/N_{\pi+} /10^{-2}$ & $3.79 \pm 0.69 \pm$ & $3.90 \pm 0.28 \pm$ & $3.66 \pm 0.19 \pm$ & $4.39 \pm 0.18 \pm$ & $4.11 \pm 0.18 \pm$ \\
 Measured range & $1.52$ & $0.55$ & $0.53$ & $0.51$ & $0.68$ \\
 & & & & & \\
$N_{K+}^{tot}/N_{\pi+}^{tot}$ & $3.27 \pm 0.6 \pm$ & $3.16 \pm 0.23 \pm$ & $2.75 \pm 0.14 \pm$ & $ 3.16 \pm 0.13 \pm$ & $ 2.83 \pm 0.12 \pm$ \\
$/10^{-2}$, Full kin. & $1.38 \pm 0.79$ & $0.54 \pm 0.71$ & $0.48 \pm 0.54$ & $0.48 \pm 0.52$ & $0.54 \pm 0.39$ \\
range & & & & & \\
 & & & & & \\
$K^+$ inv. slope $T_0$, &  $67 \pm 12 \pm 12$        & $80\pm 7\pm 5 $    & $81\pm 5\pm 5$   & $81\pm 5\pm 4$     & $78\pm 5\pm 4$ \\
MeV, Meas. range & & & & &\\
\hline
\end{tabular}
\end{footnotesize}
\label{yield_tablepiK}
\end{table}

The measured $\pi^+$ and $K^+$ meson multiplicities are extrapolated to the entire kinematic
range using the averaged extrapolation factors obtained from the predictions of the DCM-SMM,
UrQMD and PHSD models shown in Table~\ref{extrap_tablepiK}.
The largest difference of the extrapolation factors from their average value is taken
as systematic uncertainty of the extrapolation factor.

\begin{table}[!hbp]
  \caption {Extrapolation factors for $\pi^+$ and $K^+$ multiplicities, from the measured
  range to the entire kinematical range, obtained as an average of the extrapolation
  factors derived from the DCM-SMM, PHSD, and UrQMD models. The maximum difference
  between the model factors from their averaged value is taken as the uncertainty of the
  extrapolation factors.   $A_{\mathrm{part}}$ is the number of participant nucleons obtained
  from the DCM-SMM model.
 $\sigma_{inel}$ is the inclusive cross section for inelastic Ar+A interactions.}
\begin{footnotesize}
\vspace{0.3cm}
\begin{tabular}{|l|c|c|c|c|c|}
\hline
& & & & & \\
& Ar+C & Ar+Al & Ar+Cu & Ar+Sn & Ar+Pb \\
& & & & & \\
\hline
& & & & & \\
$\pi^+$ Extrap. factor & $3.25 \pm 0.18$ & $3.73 \pm 0.13$ & $4.45 \pm 0.07$ & $5.12 \pm 0.26$ & $5.91 \pm 0.55$ \\
 & & & & & \\
$K^+$ Extrap. factor & $2.81 \pm 0.66$ & $3.02 \pm 0.67$ & $3.34 \pm 0.65$ & $3.7 \pm 0.58$ & $4.1 \pm 0.43$  \\
 & & & & & \\
$A_{\mathrm{part}}$, {\tiny DCM-SMM}                &   14.8                                    &  23.0                             &  33.6       &  48.3                                  &  63.6 \\
 & & & & & \\
$\sigma_{inel}$, mb  \cite{HadesL0}                              & $1470 \pm 50$  & $1860 \pm 50$ & $2480 \pm 50$ & $3140 \pm 50$  & $3940 \pm 50$  \\
& & & & & \\
\hline
\end{tabular}
\end{footnotesize}
\label{extrap_tablepiK}
\end{table}

 The multiplicities of $K^+$ and $\pi^+$ mesons and their  ratios
are summarized in Table~\ref{yield_tablepiK}. The $K^+$ to $\pi^+$  ratios do not
show a significant dependence on the mean number of participant nucleons
$A_{\mathrm{part}}$,
 determined as  the mean number of nucleons that underwent at
 least one inelastic collision.
 The values of $A_{\mathrm{part}}$, based on the DCM-SMM model,
are listed in Table~\ref{extrap_tablepiK}.

\begin{figure}[tbh]
\begin{center}
\vspace{-0.5cm}
\includegraphics[width=0.95\textwidth,bb=0 0 822 679]{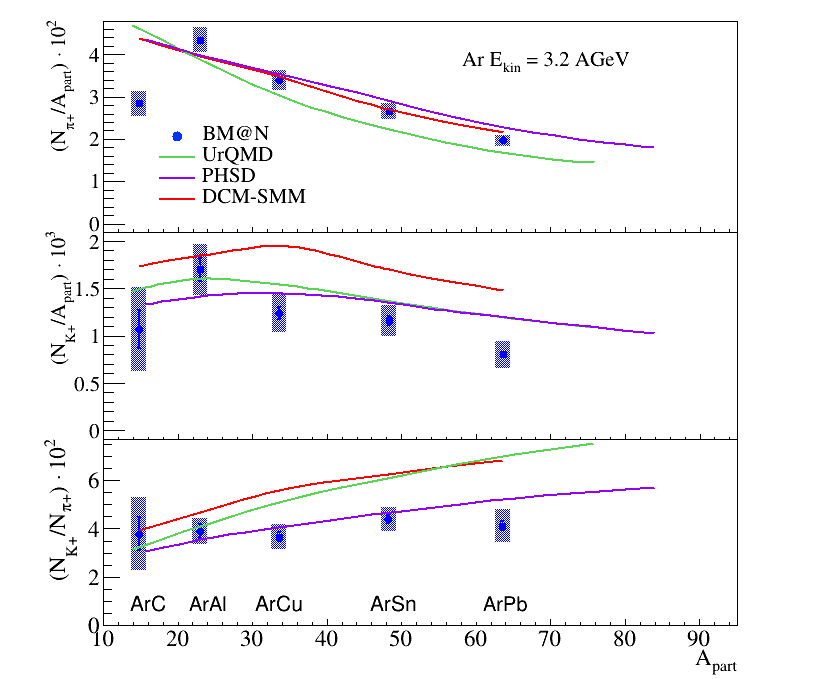}
 \end{center}
\vspace{-0.8cm}
\caption{Ratios of the $\pi^+$ (a) and $K^+$ (b) multiplicities to the number of participant
nucleons and ratios of the $K^+$ to $\pi^+$ multiplicities (c) in the measured kinematic range
in Ar+C, Al, Cu, Sn, Pb interactions. The vertical bars and boxes represent the statistical
and systematic uncertainties, respectively.
The results are compared with predictions of the DCM-QGSM, UrQMD and PHSD models for
argon-nucleus interactions shown as red, green, and magenta lines.}
\label{piyield_npart}

\vspace{-12.3cm}
\hspace{12.5cm} (a)

\vspace{2.9cm}
\hspace{12.5cm} (b)

\vspace{2.4cm}
\hspace{12.5cm} (c)
\vspace{5.5cm}

\end{figure}

The $\pi^+$ and $K^+$ multiplicities per participant nucleon, $A_{\mathrm{part}}$,
are plotted and compared to predictions of the DCM-SMM, UrQMD and PHSD models in
 Figs.~\ref{piyield_npart}.  For $\pi^+$, the three models predict a steady decrease
 of this ratio with increasing atomic weight of the target, from C to Pb. This behavior
 is observed in the data with all targets with the exception of the C target.
A similar trend  is also observed in the data for ratios of the $K^+$
multiplicities to $A_{\mathrm{part}}$.
The $K^+$ to $\pi^+$ multiplicity ratios are shown in Fig.~\ref{piyield_npart}c.
They show no dependence on the number of participant nucleons contrary to the three
models that show a small increase. The PHSD model exhibits a rather small increase
and, within the experimental uncertainties, is compatible with the measured $K^+$
to $\pi^+$ multiplicity ratios.

The $\pi^+$ and $K^+$ meson multiplicities  in argon-nucleus interactions can be
compared with previously published results.
The HADES experiment measured Ar+KCl interactions at the lower beam kinetic energy
of 1.76 AGeV \cite{HADES1,HADES2,HADES3}. The total $\pi^-$ and $K^+$ multiplicities
in semi-central events (with an average number of participant nucleons $A_{\mathrm{part}}$
of 38.5) was reported to be 3.9 and $2.8 \cdot 10^{-2}$, respectively. The results presented
here for Ar+Cu interactions at the beam kinetic energy of 3.2 AGeV  ($A_{\mathrm{part}}$ of 33.6,
see Table~\ref{extrap_tablepiK}) are higher by factors of 1.3 and 5 relative to the HADES
results. The difference in the $K^+$ multiplicities could be explained by the energy
dependence of the  $K^+$ cross section near the kinematic threshold for $K^+$
production ($E_{thr}(NN)\sim 1.58$ GeV).
The effective inverse slope parameters obtained by HADES
from the $m_T$ spectra of $\pi^-$ and $K^+$ extrapolated to $y^* = 0$ are 82.4 MeV
and 89 MeV, respectively, comparable to those reported here
(see Figs.~\ref{T0pi_rapidity} and \ref{T0K_rapidity}).

The FOPI experiment measured Ni+Ni interactions at the beam kinetic energy
of 1.93 AGeV \cite{FOPI1,FOPI2,FOPI3}. Consistent results were also reported by
the KaoS experiment that measured the $K^+$ multiplicities in Ni+Ni interactions
at kinetic energies of 1.5 and 1.93 AGeV \cite{KaoS1,KaoS2}.
 The total $K^+$ multiplicity in
 semi-central and central Ni+Ni interactions, with $A_{\mathrm{part}}$ of 46.5 and 75,
 were reported to be $3.6\cdot 10^{-2}$ and $8.25\cdot 10^{-2}$, respectively.
 These values can be compared with the results reported here in Table~\ref{yield_tablepiK}
 for the various targets. The $K^+/\pi^+$ multiplicity ratio measured by FOPI in triggered semi-central
 events is $7.6 \cdot 10^{-3}$, which is by a factor $\sim$ 4 smaller than the $K^+/\pi^+$ multiplicity
 ratio obtained here in Ar + Sn interactions for the entire kinematical range
 ($A_{\mathrm{part}}$ of 48.3, see Table~\ref{extrap_tablepiK}).
 It should be taken into account that the beam kinetic energy of the FOPI experiment (1.93 AGeV) is
 lower than that of the BM@N experiment. The effective inverse slope of 110.9 MeV, estimated by
 FOPI at $y^* = 0$ from the $K^+$ transverse mass spectrum  is consistent within uncertainties with
 the inverse slope parameter $T_0$, reported here for $K^+$ in the range $y^* \gtrsim 0$
 (see Fig. \ref{T0K_rapidity}). The consistency of the inverse slope parameters reported here
 with the results of the HADES and FOPI experiments indicates the absence of a strong
 dependence of $T_0$ on the beam energy and atomic weight of the colliding nuclei.

\begin{figure}[tbh]
\begin{center}
\vspace{0.0cm}
\includegraphics[width=0.9\textwidth]{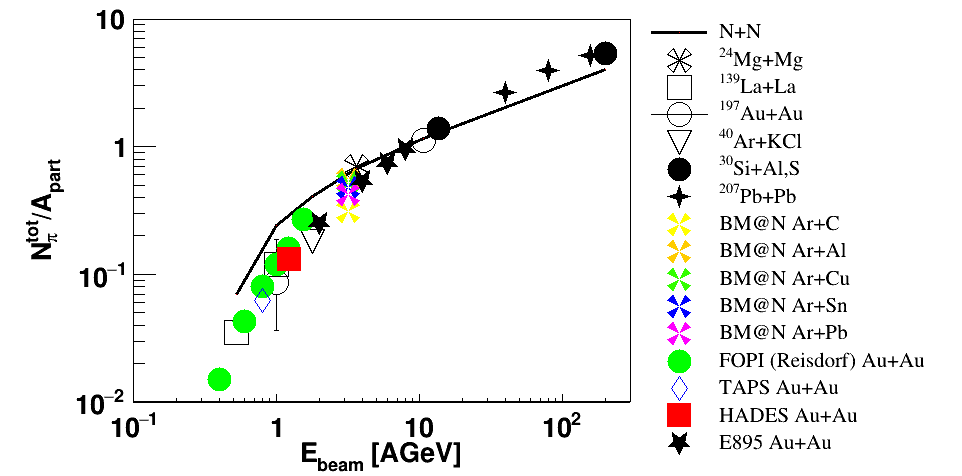}
\end{center}
\vspace{-0.4cm}
\caption{Comparison of the BM@N results to the world measurements (references in the text)
of the total pion multiplicity $N_{\pi}^{tot}$ per participant nucleon
	 $A_{\mathrm{part}}$ as a function of the beam kinetic energy $E_{\rm{beam}}$.}
\label{npi_ebeam}
\end{figure}

\begin{figure}[tbh]
\begin{center}
\vspace{0.0cm}
\includegraphics[width=0.9\textwidth]{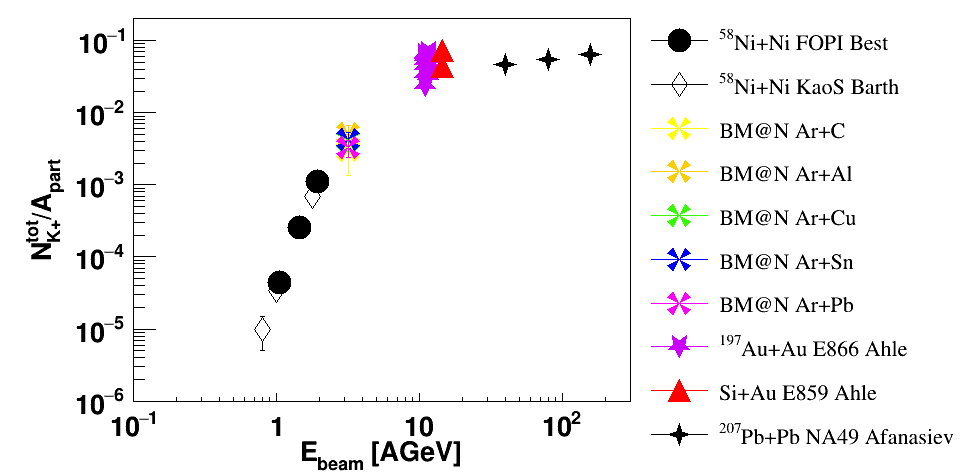}
\end{center}
\vspace{-0.4cm}
\caption{Comparison of the BM@N results to the world measurements (references in the text)
         of the $K^+$ multiplicity per participant nucleon
	 $A_{\mathrm{part}}$ as a function of the beam kinetic energy $E_{\rm{beam}}$.}
\label{nK_ebeam}
\end{figure}

The total pion multiplicity $N_{\pi}^{tot}$, where $N_{\pi}^{tot}=N_{\pi^+}^{tot}+N_{\pi^-}^{tot}+N_{\pi^0}^{tot}$,
normalized to the average number of participant nucleons $A_{\mathrm{part}}$ are compiled
in  Fig.~\ref{npi_ebeam} for different collision systems and beam energies.
References~\cite{Senger_1999} and ~\cite{Adamczewski_HADES_2020} contain a compilation
of pion data for interactions of nucleon-nucleon  (N+N)~\cite{Gazdzicki_NN_1995},
Mg+Mg~\cite{Anikina_MgMg_Nuclotron_1989}, La+La~\cite{Harris_LaLa_Bevalac_1987}, Au+Au~\cite{Pelte_AuAu_FOPI_1997,Wagner_AuAu_KaoS_1998,Schwalb_AuAu_TAPS_1994},
 Ar+KCl~\cite{Harris_ArKCl_Bevalac_1985}, Si+Al, S+S~\cite{Abbott_SiAlSS_AGS_E802_1993,Bachler_SiAlSS_AGS_E810_1994},
 Pb+Pb ~\cite{Jacobs_PbPb_NA49_1997,Afanasiev_PbPb_NA49_2002},
 Au+Au~\cite{Reisdorf_AuAu_FOPI_2007, Wolf_AuAu_TAPS_1998,Averbeck_AuAu_TAPS_2003, Klay_AuAu_AGS_E895_2003}.
 To estimate $N_{\pi}^{tot}$
from the $\pi^+$ multiplicities reported here, the predictions  of the DCM-SMM model are used.
 The total $K^+$ multiplicity in the entire kinematic range normalized to the average number of
 participant  nucleons $A_{\mathrm{part}}$ are compiled in  Fig.~\ref{nK_ebeam}. The figure
 includes the world data taken from ~\cite{FOPI1,Barth_NiNi_KaoS_1997,Ahle_AuAu_SiAu_AGS_E802_1999,
 Ahle_AuAu_AGS_E802_1998,Afanasiev_PbPb_NA49_2002} together with the results reported here.
Figures~\ref{npi_ebeam} and~\ref{nK_ebeam} demonstrate that the BM@N results reported here are
consistent with the world data on the production of $\pi$ and $K^+$ mesons.

\section{Summary}
\label{sect7}

First physics results of the BM@N experiment are presented on the $\pi^+$ and $K^+$ meson yields and
their ratios in argon-nucleus interactions at the beam kinetic
 energy of 3.2~AGeV. The results are compared with the DCM-SMM, UrQMD and PHSD models and with
 the previously published results of other experiments.

The  inverse slope parameter $T_0$ of the $\pi^+$ transverse momentum spectrum is about 40 MeV
in the forward rapidity range, rising to 90 MeV in the central rapidity range. In general,
the y-dependence of $T_0$  is consistent with the predictions of the models, but there is a
tendency for the experimental results to show a flatter dependence of the slope values in the
central rapidity range compared to a rising dependence predicted by the models.

The $T_0$ value for $K^+$ exhibits a weak dependence on the rapidity. The PHSD and DCM-SMM
models reproduce the weak dependence of $T_0$, whereas UrQMD predicts much larger $T_0$ values.

The ratios of the $K^+$ to $\pi^+$ multiplicities show no significant dependence on the mean
number of participant nucleons $A_{\mathrm{part}}$ in argon-nucleus collisions.
The PHSD prediction is compatible with this result, whereas the DCM-SMM and UrQMD models predict
a smooth rising of the $K^+$ to $\pi^+$ ratio with $A_{\mathrm{part}}$.

The $\pi$ and $K^+$ multiplicities normalized to $A_{\mathrm {part}}$ are found to be consistent
with the rising energy  dependence of the world data on the production of $\pi$ and $K^+$ mesons
measured for various colliding nuclei and beam energies.

\paragraph{Acknowledgments.}
The BM@N Collaboration acknowledges the efforts of the staff of the accelerator division of the
Laboratory of High Energy Physics at JINR that made this experiment possible, I.Tserruya
and V.Pozdniakov for fruitful discussions of the analysis and results.
The BM@N Collaboration acknowledges support of the HybriLIT of JINR, HPC Village project and HGPU
group for the provided computational resources.

%
%

\clearpage

\vspace{0.0cm}

\begin{thebibliography}{99}

\bibitem{Friman}
 B.~Friman, W.~N$\ddot{o}$renberg, and V.D.~Toneev, Eur. Phys. J. A 3 (1998).

\bibitem{Cleymans} J.~Randrup and J.~Cleymans, Phys. Rev. C 74 (2006) 047901.

\bibitem{NICAWhitePaper} NICA White Paper, Eur. Phys. J. A 52 (2016).


\bibitem{BMN_CDR} BM@N Conceptual Design Report:
                  \url{http://nica.jinr.ru/files/BM@N/BMN_CDR.pdf}

\bibitem{Fuchs} Ch.~Fuchs, Prog. Part. Nucl. Phys. 56 (2006) 1-103.
\bibitem{BMN_QM} M.~Kapishin (for the BM@N Collaboration), Nucl. Phys. A 982 (2019) 967-970.

\bibitem{BMN_SQM} M.~Kapishin (for the BM@N Collaboration), SQM 2019 proceedings,
                  285 Springer Proc. Phys. 250 (2020) 21-27.

\bibitem{BMN_project} BM@N project:

\url{http://nica.jinr.ru/files/BM@N/BMN_project.pdf}

\bibitem{BMN_GEM} D.~Baranov et al., JINST 12 (2017) no. 06, C06041

\bibitem {KA_2022} \textcolor{black}{K.~Alishina et al., Phys. Part. Nucl., 53 (2022) no. 2, 470-475.}

\bibitem {BMN_ToF400} \textcolor{black}{V. Babkin et al., Nucl. Instrum. Meth. A 824, P.490-492 (2016);
         V. Babkin et al., Proceedings of Science, 2014, Vol.213 (Proceedings of TIPP-2014), P.289}.

\bibitem {BMN_ToF700} \textcolor{black}{N. Kuzmin et al., Nucl. Instrum. Meth. A 916, P. 190-194 (2019)}.

\bibitem{Kisel}
V.~Akishina and I.~Kisel, J. Phys.: Conf. Ser. 599, 012024 (2015),
I.~Kisel, Nucl. Instrum. Meth. A 566, 85 (2006).

\bibitem{DCM_QGSM} N.~Amelin, K.~Gudima, and V.~Toneev, Sov. J. Nucl. Phys. 51, 1093 (1990).

\bibitem{DCM_SMM} M. Baznat, A. Botvina, G. Musulmanbekov, V. Toneev, V. Zhezher,  Phys. Part. Nucl. Lett. 17 (2020) no. 3; arXiv: 1912.09277v.

\bibitem{GEANT3} CERN Program Library, Long Writeup W5013, Geneva, CERN, 1993.

\bibitem{BmnRoot} \url {https://git.jinr.ru/nica/bmnroot}

\bibitem{Garfield} \url{http://garfieldpp.web.cern.ch/garfieldpp}

\bibitem {HadesL0} K.~Kanaki, PhD Thesis, Technische Universit$\ddot{a}$t Dresden, 2007.

\bibitem {AngelovCC} H.~Angelov et al., P1-80-473, JINR, Dubna.

\bibitem{UrQMD} S. A.~Bass et al., Prog. Part. Nucl. Phys. 41 225 (1998).

\bibitem{PHSD} W.~Cassing and E. L.~Bratkovskaya, Nucl. Phys. A 831 (2009) 215-242.

\bibitem {HADES1} G.~Agakishiev et al., HADES Collaboration, Eur. Phys. J. A 47 (2011) 21.

\bibitem {HADES2} G.~Agakishiev et al., HADES Collaboration, Phys. Rev. C 80 (2009) 025209.

\bibitem {HADES3} G.~Agakishiev et al., HADES Collaboration, Phys. Rev. C 82 (2010) 044907.

\bibitem {FOPI1} D.~Best et al., FOPI Collaboration, Nucl. Phys. A 625 (1997) 307-324.

\bibitem {FOPI2} N.~Bastid et al., FOPI Collaboration, Phys. Rev. C 76 (2007) 024906.

\bibitem {FOPI3} K.~Piasecki et al., FOPI Collaboration, Phys. Rev. C 99 (2019) 1, 014904.

\bibitem {KaoS1} M.~Menzel et al., KaoS Collaboration,  Phys. Lett. B 495 (2000) 26-32.

\bibitem {KaoS2} A.~Forster et al., KaoS Collaboration,  Phys. Rev. C 75 (2007) 024906.

\bibitem {Senger_1999} \textcolor{black}{P.~Senger et al., J. Phys. G 25 (1999) R59-R131.}

\bibitem {Adamczewski_HADES_2020} \textcolor{black}{J. Adamczewski-Musch et al., Eur. Phys. J. A 56 (2020) 259.}

\bibitem {Gazdzicki_NN_1995} Gazdzicki M. and R$\ddot{o}$hrich D., 1995 Z. Phys. C 65 215.
\bibitem {Anikina_MgMg_Nuclotron_1989} Anikina et al., JINR Rapid Comm Dubna, 1 (1989) 12.
\bibitem {Harris_LaLa_Bevalac_1987} Harris J. W. et al., 1987 Phys. Rev. Lett. 58 463.
\bibitem {Pelte_AuAu_FOPI_1997} Pelte D. et al., 1997 Z. Phys. A 357 215.
\bibitem {Wagner_AuAu_KaoS_1998} Wagner A. et al., 1998 Phys. Lett. B 420 20.
\bibitem {Schwalb_AuAu_TAPS_1994} Schwalb O. et al., 1994 Phys. Lett. B 321 20.
\bibitem {Harris_ArKCl_Bevalac_1985} Harris J. W. et al., 1985 Phys. Lett. B 153 377.
\bibitem {Abbott_SiAlSS_AGS_E802_1993} T. Abbott et al. (E-802 Collaboration), Phys. Rev. C 50 (1993) 1024.
\bibitem{Bachler_SiAlSS_AGS_E810_1994} J. Bachler et al., Phys. Rev. Lett. 72 (1994) 1419.
\bibitem {Jacobs_PbPb_NA49_1997} Jacobs P. and NA49 Collaboration 1997 Proc. of the 3rd Int. Conf. on the
           Physics and Astrophysics of the Quark Gluon Plasma (Jaipur, India) (Delhi: Narosa).

\bibitem{Afanasiev_PbPb_NA49_2002} Afanasiev S. V. et al., Phys. Rev. C. 66 054902 (2002).

\bibitem {Reisdorf_AuAu_FOPI_2007} W. Reisdorf et al. (FOPI Collaboration), Nucl. Phys. A 781, 459 (2007).
\bibitem {Wolf_AuAu_TAPS_1998} A.R. Wolf et al. (TAPS Collaboration), Phys. Rev. Lett. 80, 5281 (1998).
\bibitem {Averbeck_AuAu_TAPS_2003} R. Averbeck et al., Phys. Rev. C 67, 024903 (2003).
\bibitem {Klay_AuAu_AGS_E895_2003} J.L. Klay et al. (E895 Collaboration), Phys. Rev. C 68, 054905 (2003).

\bibitem{Barth_NiNi_KaoS_1997} R. Barth et al. Phys. Rev. Lett. 78 (1997), p. 4007.
\bibitem{Ahle_AuAu_SiAu_AGS_E802_1999} L. Ahle et al. (E802 Collaboration), Phys. Rev. C 60, 044904 (1999).
\bibitem{Ahle_AuAu_AGS_E802_1998} L. Ahle et al. (E802 Collaboration), Phys. Rev. C 58, 3523 (1998).

\end{thebibliography}
\end{document}